\documentclass{aa}
\usepackage{graphicx}
\usepackage{amsmath, amssymb}
\usepackage[utf8]{inputenc}
\usepackage{txfonts}
\usepackage{array}
\usepackage{tikz}
\usepackage{hyperref}

\hypersetup{
    colorlinks = true,
    allcolors = {blue}
}
\definecolor{lime}{HTML}{A6CE39}
\DeclareRobustCommand{\orcidicon}{%
  \begin{tikzpicture}[baseline=-0.55ex]
    \filldraw[lime] (0,0) circle[radius=0.145];
    \node[white] at (0,0) {\fontfamily{phv}\selectfont\tiny i\scalebox{0.78}{D}};
  \end{tikzpicture}%
}
\newcommand{\orcidmark}[1]{\hspace{0.14em}\href{https://orcid.org/#1}{\orcidicon}}

\makeatletter
\newlength{\aa@sidecapfight}
\def\sidecaption#1\caption{\setbox0=\hbox{\ignorespaces#1\unskip}\aa@sidecaptionwidth=\linewidth \advance\aa@sidecaptionwidth -\wd0 \advance\aa@sidecaptionwidth -\aa@figgap \ifdim\aa@sidecaptionwidth<\aa@sidecaptionminwidth \let\aa@caption\@caption \unhbox0 \else \let\aa@caption\aa@sidecaption \dimen0=\dp0 \advance\dimen0 by \ht0 \dp0=\dp\strutbox \advance\dimen0 by -\dp0 \ht0=\dimen0 \global\aa@sidecapfight=\ht0 \unhbox0 \hspace{\aa@figgap}\fi \refstepcounter\@captype \@dblarg{\aa@caption\@captype}}
\long\def\aa@makesidecaption#1#2{\setbox\@tempboxa=\hbox{\parbox[b]{\aa@sidecaptionwidth}{\aa@captionfont\aa@@makecaption{#1}\aa@capstrut #2}}\dimen@=\aa@sidecapfight \advance\dimen@ by -\ht\@tempboxa \ifdim\dimen@<\z@ \dimen@=\z@ \fi \raise\dimen@\box\@tempboxa}
\makeatother

\titlerunning{Host-star metallicities and kinematics of directly imaged brown-dwarf companions}
\authorrunning{Swastik et al.}
\begin{document}

\renewcommand{\today}{15 September 2026}

   \title{Host-star metallicities and kinematics of directly imaged brown-dwarf companions}

\author{
C.~Swastik\inst{\ref{unimi}}\thanks{Corresponding author: \href{mailto:swastik.chowbay@unimi.it}{\texttt{swastik.chowbay@unimi.it}}}\orcidmark{0000-0003-1371-8890}
\and R.~K.~Banyal\inst{\ref{iia}}\orcidmark{0000-0003-0799-969X}
\and M.~Muduli\inst{\ref{vtu}}\orcidmark{0009-0003-2810-2551}
\and S.~Soni\inst{\ref{vssc}}\orcidmark{0009-0008-3492-9180}
\and A.~Jino\inst{\ref{stalberts}}\orcidmark{0009-0001-1165-657X}
\and S.~Facchini\inst{\ref{unimi}}\orcidmark{0000-0003-4689-2684}
\and Z.~Wahhaj\inst{\ref{inst-eso}}\orcidmark{0000-0001-8269-324X}
\and G.~Lodato\inst{\ref{unimi}}\orcidmark{0000-0002-2357-7692}
\and P.~Saraf\inst{\ref{prl}}\orcidmark{0009-0001-4813-0432}
\and A.~Choudhary\inst{\ref{ism}}\orcidmark{0009-0004-6657-2260}
\and A.~K.~Bhavya\inst{\ref{puc}}
\and M.~P.~Navaneeth\inst{\ref{iia}}
\and B.~Banerjee\inst{\ref{iia}}\orcidmark{0000-0001-8075-3819}
\and S.~Biswas\inst{\ref{iia}}\orcidmark{0009-0000-2401-9986}
\and T.~Sivarani\inst{\ref{iia}}\orcidmark{0000-0003-0891-8994}
\and G.~Maheswar\inst{\ref{iia}}\orcidmark{0009-0007-0745-9147}
\and A.~Surya\inst{\ref{iia}}\orcidmark{0000-0002-9967-0391}
}
\institute{
Dipartimento di Fisica, Universit\`a degli Studi di Milano, Via Celoria 16, 20133 Milano, Italy \label{unimi}
\and Indian Institute of Astrophysics, Koramangala 2nd Block, Bengaluru 560034, Karnataka, India \label{iia}
\and Visvesvaraya Technological University, Belagavi 590018, Karnataka, India \label{vtu}
\and Space Physics Laboratory, Vikram Sarabhai Space Centre, ISRO \label{vssc}
\and St.~Albert's College, Ernakulam 682018, Kerala, India \label{stalberts}
\and European Southern Observatory, Alonso de C\'ordova 3107, Vitacura Casilla 19001, Santiago, Chile \label{inst-eso}
\and Physical Research Laboratory, Navrangpura, Ahmedabad 380009, Gujarat, India \label{prl}
\and Indian School of Mines, Indian Institute of Technology, Dhanbad 826004, Jharkhand, India \label{ism}
\and Instituto de Astrof\'isica, Pontificia Universidad Cat\'olica de Chile, Vicu\~na Mackenna 4860, Macul, Santiago, Chile \label{puc}
}

   \date{Received 2 June 2026 / Accepted 15 September 2026}

\abstract
{Brown dwarfs are common as free-floating objects but rare as close companions to Sun-like stars, a disparity known as the ``brown-dwarf desert.'' Host-star metallicity can constrain whether these companions form mainly through metal-sensitive core accretion or through less metal-dependent disk or cloud fragmentation.}
{We extend our homogeneous spectroscopic analysis of directly imaged planet hosts into the brown-dwarf regime and compare their metallicities with those of planet hosts and close-in brown-dwarf hosts.}
{We compiled 54 unique directly imaged brown-dwarf systems selected over an inclusive $13$--$80~M_{\rm Jup}$ interval and projected separations from about 5~au to several thousand au. Objects near the model-dependent $70$--$75~M_{\rm Jup}$ hydrogen-burning boundary may, however, be very low-mass stars. For 31 hosts with archival high-resolution spectra, we derived atmospheric parameters and metallicities using Bayesian spectral synthesis. Literature companion masses and projected separations are heterogeneous and often model-dependent, and are used only for demographic context. Galactic velocities were calculated for 46 hosts solely to characterize the youth-biased imaging sample.}
{The host stars have a broadly solar metallicity distribution, with a median [Fe/H] of $+0.06$~dex and a median absolute deviation of $0.11$~dex, and show no strong metal-rich bias. No statistically significant metallicity difference is detected between the  directly imaged  lower- and higher-mass subsamples at the current sample size. The hosts are kinematically cold, as expected from the youth-biased selection of direct-imaging surveys.}
{The absence of a strong metal-rich bias suggests that classical core accretion does not dominate the wide-orbit brown-dwarf population. Disk instability and cloud fragmentation remain plausible, but the current sample and heterogeneous companion properties do not permit object-by-object discrimination between these channels. The comparisons involving the smallest samples remain exploratory.}

   \maketitle

\section{Introduction}

\label{sec:intro}

Brown dwarfs occupy a unique position in the substellar mass spectrum. With masses of approximately $13$--$75~M_{\rm Jup}$, they are bracketed by the deuterium-burning and hydrogen-burning limits. Furthermore, they are well established as a numerous free-floating population in star-forming regions and the field \citep{Rebolo1995}. As companions to Sun-like stars, however, they are markedly underrepresented at close orbital separations relative to both giant planets (GPs) and stellar binaries, a deficit referred to as the ``brown-dwarf desert'' \citep[e.g.,][]{Marcy2000,2000A&A...355..581H,Grether2006}. The desert points to a transition in formation and evolutionary pathways near the planet--brown-dwarf boundary, and identifying its physical origin remains a central question in substellar demographics. Unlike stars, brown dwarfs do not sustain core hydrogen fusion, although they can fuse deuterium and, in some cases ($\gtrsim 65~M_{\rm Jup}$), lithium \citep{Burrows1997,2000ARA&A..38..485B,Chabrier2000,Burrows2001,Spiegel2011,2023A&A...671A.119C}. Their intrinsic faintness at optical wavelengths, combined with their predominantly infrared emission, makes brown dwarfs challenging to detect as stellar companions \citep{Goldman1999,Bowler2016,Kirkpatrick2021}. The first unambiguous brown-dwarf companion, GJ~229~B, was confirmed only in 1995 through direct imaging \citep{Nakajima1995,Oppenheimer1995}. Two decades of monitoring established its dynamical mass \citep{Brandt2020}.  Despite being a well-studied benchmark, only recently was it resolved into a close binary of two brown dwarfs, with individual masses of $38.1\pm1.0$ and $34.4\pm1.5~M_{\rm Jup}$ \citep{Xuan2024Natur}. Quantitatively, the deficit is sharpest at orbital periods $P \lesssim 100$~days and companion masses $\sim35$--$55~M_{\rm Jup}$ \citep{Grether2006,Ma2014}, where radial-velocity (RV) surveys are most sensitive. Complementary constraints at intermediate separations of $\sim 5$--$50$~au, derived from \textsc{Hipparcos}--\emph{Gaia} astrometric accelerations \citep{Brandt2021}, probe a longer-period regime inaccessible to orbital-averaged RV signals and reveal a similar paucity of intermediate-mass brown-dwarf companions, indicating that the scarcity is not primarily an observational artifact of the RV technique.

\begin{figure*}
\sidecaption
\includegraphics[width=11cm]{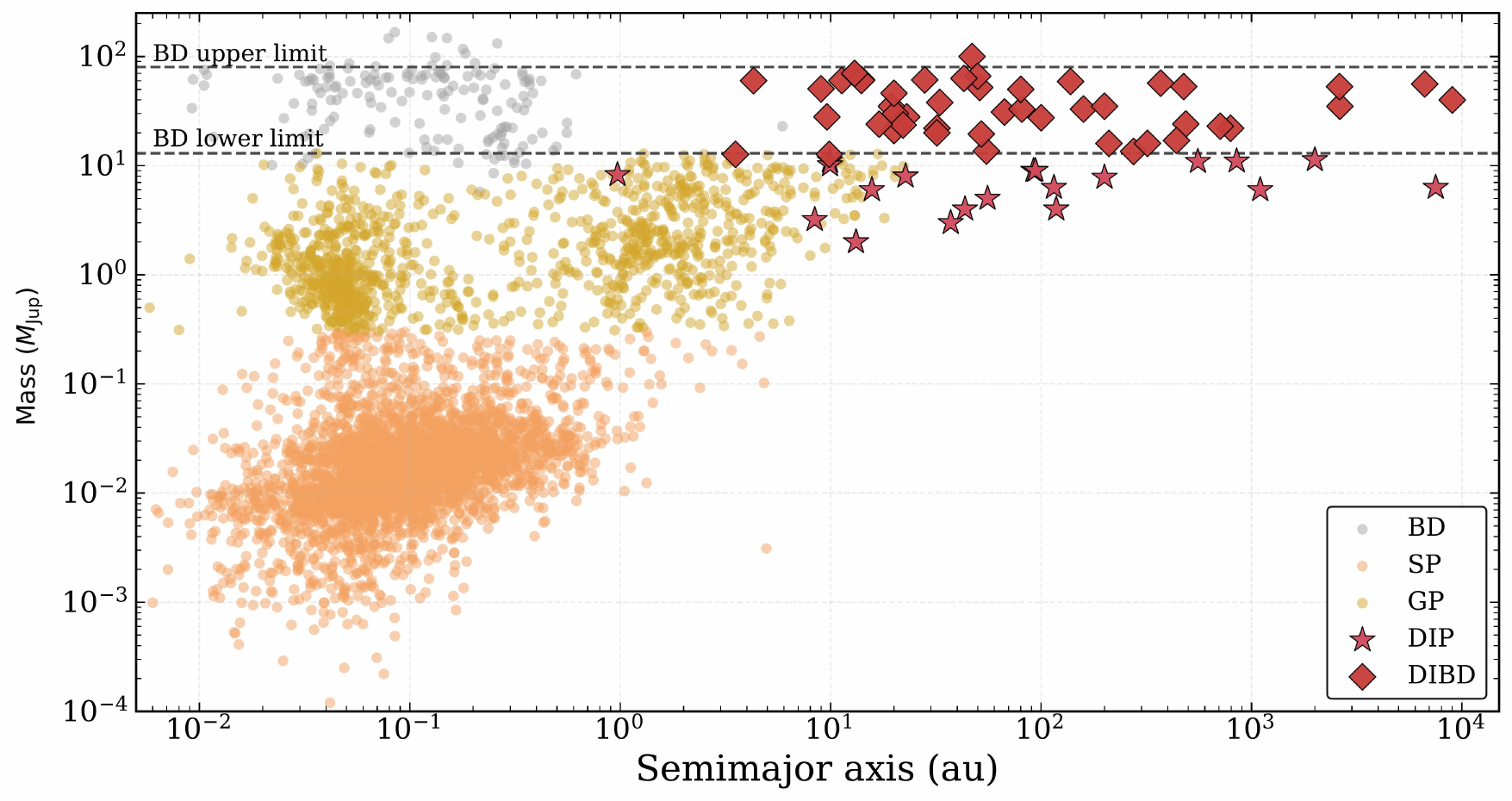}
\caption{Companion mass vs. orbital separation for substellar companions. For most directly imaged companions the projected separation is used as a proxy for the orbital semimajor axis. SPs, GPs, and literature brown dwarfs (BD) are background populations, while DIPs and DIBD candidates are highlighted. The lower and upper dashed lines mark the adopted $13~M_{\rm Jup}$ and $80~M_{\rm Jup}$ sample-selection bounds, respectively.}
\label{mass_sem}
\end{figure*}

Brown-dwarf companions on wide orbits (tens to hundreds of au) are primarily detected through high-contrast direct-imaging surveys \citep[e.g.,][]{Bowler2016,Nielsen2019,Swastik2026LkCa}. These systems probe a complementary region of parameter space that overlaps with massive planets and provides leverage on formation mechanisms that close-in samples alone cannot access. Three formation pathways are commonly discussed: (i)~molecular-cloud fragmentation, analogous to stellar binary formation \citep{Bate2002,Chabrier2014}; (ii)~gravitational instability in massive protoplanetary disks \citep{Rice2005,Stamatellos2009,Kratter2016}; and (iii)~core accretion followed by runaway gas accretion \citep{Pollack1996,Mordasini2012}. These formation mechanisms predict different host-star metallicity signatures. Core accretion is expected to favor metal-rich environments, whereas disk instability and molecular-cloud fragmentation are comparatively insensitive to metallicity and instead reflect the properties of the parent disk or molecular cloud. We used the homogeneous host-star metallicities as the primary population-level formation diagnostic in this work.

Galactic kinematics provide information on stellar population membership and are sensitive to age \citep[e.g.,][]{Banerjee2024,2024AJ....167..270S}. However, in directly imaged samples they are strongly shaped by the youth bias of imaging target selection (Sect.~\ref{sec:results_kinematics}). We consequently used the kinematics only to characterize the selected population and to confirm its youth, not as an independent diagnostic of the companion-formation channel. Observationally, a transition near $\sim40$--$45~M_{\rm Jup}$ has been suggested
in several diagnostics---the companion mass function \citep{Sahlmann2011}, the
orbital eccentricity distribution \citep{Ma2014}, and the host-star metallicity
distribution \citep{Mata2014,Maldonado2017}---with lower-mass brown dwarfs
showing planet-like characteristics and higher-mass objects resembling stellar
companions. A related structural transition near $\sim60~M_{\rm Jup}$ was identified by \citet{HatzesRauer2015} from the mass--density relation, below which GPs and brown dwarfs form a single continuous sequence. More recently, the combination of precision astrometry from \textsc{Hipparcos} and \emph{Gaia} with high-contrast imaging has enabled the discovery and dynamical mass measurement of a growing number of benchmark brown-dwarf companions at separations of $\sim5$--$50$~au \citep[e.g.,][]{Brandt2021,Li2023,Franson2023}, substantially expanding the accessible parameter space.

Host-star metallicity is a key empirical diagnostic of companion formation. Giant planets preferentially orbit metal-rich stars, consistent with core-accretion models \citep{Fischer2005,Mordasini2012}, whereas stars hosting only low-mass planets show little or no metallicity enhancement \citep{Sousa2011,Buchhave2012,2018AJ....156..221N,2022AJ....164...60S}. For brown-dwarf hosts, previous studies have reported a broad distribution centered near solar metallicity with only weak or absent trends \citep{Ma2014,Maldonado2017}. Whether any mass-dependent metallicity structure exists within the brown-dwarf regime, and whether such trends extend from close-in RV samples to the wide-orbit, directly imaged regime, remain open questions that the present sample is well placed to address.

\citet{2021AJ....161..114S} presented a homogeneous spectroscopic analysis of host stars with directly imaged planetary companions. More recently, \citet{Swastik2026WISPIT2} characterized the protoplanet host WISPIT~2 through optical spectroscopy, deriving atmospheric parameters and a low-resolution, model-dependent global metallicity. The brown-dwarf regime, by contrast, has remained less uniformly characterized: a significant fraction of directly imaged brown-dwarf (DIBD) host stars have only heterogeneous or indirect metallicity estimates, which limits our ability to place them in a consistent demographic context and to compare them directly with planet hosts and with the close-in brown-dwarf population cataloged by \citet{Stevenson2023}.

In this work, we extend the homogeneous spectroscopic framework of \citet{2021AJ....161..114S} into the brown-dwarf regime. We compiled an updated sample of 54 DIBD companions ($13$--$80~M_{\rm Jup}$), spanning projected separations from $\sim 5$~au to several thousand au, and derived stellar atmospheric parameters, including effective temperatures, surface gravities, metallicities, and projected rotational velocities, for the 31 host stars with archival high-resolution spectra from HARPS, UVES, FEROS, and HIRES. Combined with \emph{Gaia}~DR3 astrometry and RVs, we examined the host-star metallicity distribution, the mass--separation demographics, and the Galactic kinematics, and compared them with directly imaged planet (DIP) hosts and with close-in brown-dwarf companions from RV and transit surveys. The homogeneous host-star metallicities are the primary new measurement of this work; the companion masses and projected separations are adopted from the literature and are used only to place the systems in the mass--separation plane. This approach allowed us to place DIBDs in a broader demographic context and to assess whether their host-star properties are consistent with distinct or overlapping formation channels across the planet--brown-dwarf boundary. The paper is organized as follows. Section~\ref{sec:sample} describes the sample selection and outlines the spectroscopic and kinematic analysis. Section~\ref{sec:results} presents the results. Section~\ref{sec:discussion} discusses their implications, and Sect.~\ref{sec:conclusions} summarizes our conclusions.

\section{Sample and analysis}
\label{sec:sample}

\subsection{Sample selection and properties}
\label{sec:sample_selection}

Our sample contains directly imaged companions selected within the nominal brown-dwarf interval $13 \le M_{\rm comp} \le 80~M_{\rm Jup}$. This range spans the interval between the deuterium-burning minimum mass \citep[$\sim13~M_{\rm Jup}$;][]{Spiegel2011,Caballero2018} and the hydrogen-burning minimum mass, which for solar composition lies near $\sim73~M_{\rm Jup}$ and ranges over $\sim70$--$75~M_{\rm Jup}$ depending on model and metallicity \citep{Chabrier2000,Burrows2001,Caballero2018}. We compiled the sample by querying the Encyclopaedia of Exoplanetary Systems \citep{Schneider2011}\footnote{\url{http://exoplanet.eu/}} and cross-referencing with recent compilations in the literature \citep[e.g.,][]{Bowler2016,Nielsen2019,Stevenson2023}. We retained only systems in which the companion has been confirmed as gravitationally bound by common proper motion, orbital monitoring, or joint RV and astrometric constraints.

Near the boundaries of the adopted mass range, we followed the classifications prevalent in the literature. We included companions whose masses are consistent with the $13~M_{\rm Jup}$ limit within their uncertainties and that are commonly treated as brown dwarfs; objects falling significantly below this threshold were classified as planetary-mass companions and excluded from the primary sample, although several were retained for comparison (Sect.~\ref{sec:comparison}). At the upper end, we included a small number of companions straddling the hydrogen-burning limit when their nominal masses remain within $80~M_{\rm Jup}$. We retained the inclusive $80~M_{\rm Jup}$ upper selection bound so as not to reject boundary objects whose mass uncertainties overlap the hydrogen-burning limit; we caution, however, that companions with nominal masses approaching or exceeding $\sim73~M_{\rm Jup}$ (most notably HR~7672~B at $75.4~M_{\rm Jup}$, with HD~19467~B and HD~4747~B also close to the limit) may be very low-mass stars rather than brown dwarfs. We treated these systems as boundary objects and avoided using their individual classifications as decisive evidence of a formation channel (Sect.~\ref{sec:disc_caveats}). For the majority of directly imaged companions, masses are derived from evolutionary models in conjunction with host-star age estimates and therefore carry uncertainties dominated by the age determination. Where model-independent dynamical masses are available from joint fits of RVs, direct-imaging astrometry, and \textsc{Hipparcos}--\emph{Gaia} proper-motion anomalies \citep[e.g.,][]{Brandt2021,Franson2022}, we adopted these in preference to model-dependent values.

\begin{figure}[t!]
\centering
\includegraphics[width=\columnwidth]{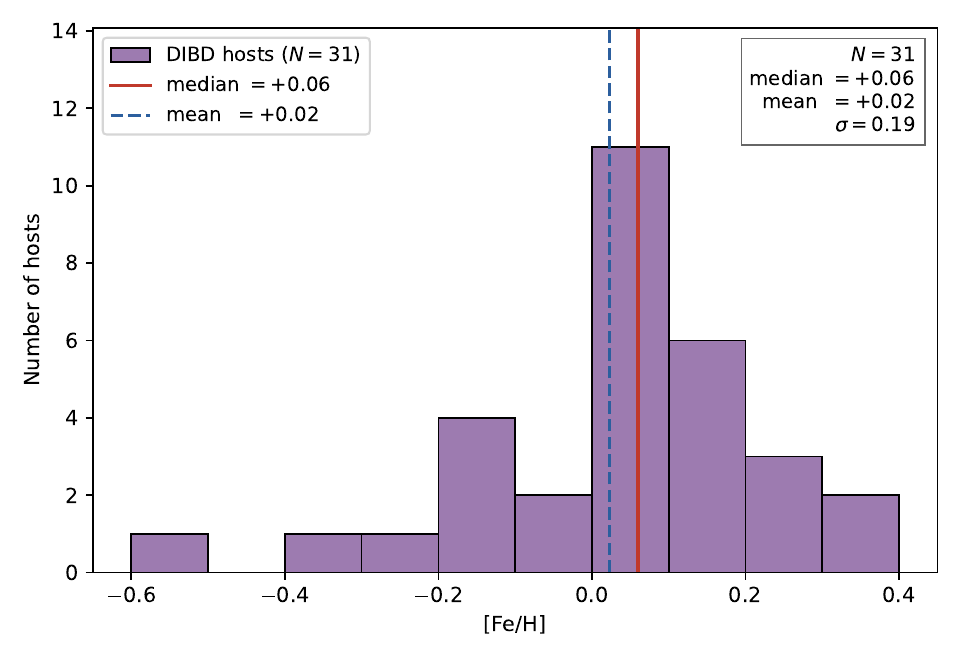}
\caption{Host-star metallicity distribution of the DIBD sample ($N=31$), binned in $0.1$~dex intervals with edges at
multiples of $0.1$~dex. The solid red and dashed~blue
lines mark the sample median ($+0.06$~dex) and mean ($+0.02$~dex),
both of which are consistent with solar; the distribution spans
$-0.6 \lesssim [\mathrm{Fe/H}] \lesssim +0.35$ with
$\sigma=0.19$~dex and shows no strong metal-rich bias.}
\label{fig:feh_hist}
\end{figure}

This procedure yielded a final sample of 54 unique DIBD systems. The sample encompasses both ``classical'' companions discovered through blind high-contrast imaging surveys \citep[e.g.,][]{Chauvin2005,Marois2008} and a growing population of benchmark brown dwarfs first identified through long-baseline RV trends or \textsc{Hipparcos}--\emph{Gaia} astrometric accelerations and subsequently confirmed via high-contrast imaging \citep[e.g.,][]{Li2023,Franson2023}. The latter category has significantly expanded the accessible parameter space toward smaller orbital separations than are typically probed by unbiased imaging surveys, linking the RV and direct-imaging detection regimes. Figure~\ref{mass_sem} shows the distribution of companion mass versus semimajor axis for the DIBD  sample in the context of the broader substellar companion population.

The host stars span spectral types from late-A to mid-M, with the majority being F-, G-, and K-type dwarfs. Many belong to young stellar associations such as the $\beta$~Pic and AB~Dor moving groups, the Scorpius--Centaurus OB association, or star-forming regions like Taurus and Chamaeleon, reflecting the well-known observational bias of direct-imaging surveys toward young systems in which substellar companions remain self-luminous and are easier to detect at wide separations. The companions span projected separations, $s = \rho\,d$, where $\rho$ is the angular separation and $d$ the distance, from $\sim 5$~au for the closest benchmark systems (e.g.,\ HD~167665~B at $5.5$~au, HD~72946~B at $6.5$~au) to several thousand au for the widest common-proper-motion pairs (e.g.,\ HD~126053~B at $\sim 2630$~au), with a median of $\sim 40$~au. For the subset of systems with a resolved orbit we quote the fitted semimajor axis $a$ in place of $s$.

\begin{table*}
\centering
\scriptsize
\renewcommand{\arraystretch}{1.3}
\caption{Stellar and companion parameters for the spectroscopically characterized directly imaged sample.}
\label{tab:star_params_all}
\begin{tabular}{l l c c c c c c c}
\hline\hline
Host star & Companion & $M_{\rm comp}$ [$M_{\rm Jup}$] & $s$ [au] & Instrument & $T_{\mathrm{eff}}$ [K] & $\log g$ & [Fe/H] (dex) & $v\sin i$ [km s$^{-1}$] \\
\hline
HD 984            & HD 984 B            & $61\pm4$                         & $28^{+7}_{-4}$              & HARPS & $6383^{+78}_{-43}$      & $4.42^{+0.16}_{-0.10}$ & $+0.10^{+0.02}_{-0.02}$ & $39.90^{+3}_{-2}$ \\
HD 19467          & HD 19467 B          & $71.6^{+5.3}_{-4.6}$             & $46.9^{+11}_{-7.4}$         & HARPS & $5681^{+24}_{-37}$      & $4.38^{+0.10}_{-0.07}$ & $-0.18^{+0.05}_{-0.04}$ & $1.31^{+0.4}_{-0.8}$ \\
HIP 74865\tablefootmark{a}   & HIP 74865 B         & $28^{+37}_{-10}$                 & $23\pm6$                    & HARPS & $6750^{+39}_{-20}$       & $4.07^{+0.15}_{-0.20}$ & $+0.07^{+0.04}_{-0.06}$ & $>75$ \\
PZ Tel            & PZ Tel B            & $28^{+25}_{-9}$                  & $27^{+14}_{-4}$             & HARPS & $5341^{+28}_{-39}$        & $4.50^{+0.07}_{-0.03}$ & $+0.00^{+0.02}_{-0.03}$ & $68.80^{+0.36}_{-0.35}$ \\
GSC 08047-00232   & GSC 08047-00232 B   & $25\pm10$                        & $264^{+81}_{-40}$           & HARPS & $4921^{+18}_{-34}$        & $4.40^{+0.03}_{-0.05}$ & $+0.03^{+0.02}_{-0.03}$ & $20.60^{+2}_{-3}$ \\
GJ 229            & GJ 229 B\tablefootmark{b}      & $71.4\pm0.6$                     & $42.9^{+3.0}_{-2.4}$        & HARPS & $3858^{+25}_{-57}$        & $4.80^{+0.06}_{-0.08}$ & $+0.17^{+0.04}_{-0.03}$ & $1.80^{+0.30}_{-0.50}$ \\
CD$-$35 2722      & CD$-$35 2722 B      & $31\pm8$                         & $204^{+48}_{-66}$           & HARPS & $3778^{+41}_{-37}$        & $4.41^{+0.02}_{-0.04}$ & $-0.15^{+0.02}_{-0.04}$ & $14.9^{+1.5}_{-1.0}$ \\
HD 206893         & HD 206893 B         & $18\pm13$                    & $8.9^{+1.4}_{-2.5}$       & HARPS & $6745^{+23}_{-34}$        & $4.62^{+0.03}_{-0.02}$ & $+0.08^{+0.03}_{-0.01}$ & $35.10^{+2}_{-3}$ \\
G 196-3           & G 196-3 B           & $15^{+30}_{-4}$                  & $\sim300$                   & HIRES & $3921^{+24}_{-10}$      & $5.02^{+0.22}_{-0.20}$ & $+0.03^{+0.03}_{-0.05}$ & $18.2^{+2.0}_{-1.5}$ \\
GJ 758            & GJ 758 B            & $37.9\pm1.5$                     & $27.4^{+5.6}_{-3.6}$        & HIRES & $5215^{+21}_{-34}$       & $4.55^{+0.03}_{-0.01}$ & $+0.06^{+0.04}_{-0.02}$ & $2.20^{+0.3}_{-0.2}$ \\
HD 3651           & HD 3651 B           & $53\pm15$                        & $\sim480$                   & HIRES & $5141^{+10}_{-21}$        & $4.65^{+0.03}_{-0.04}$ & $+0.08^{+0.02}_{-0.01}$ & $1.47^{+0.3}_{-0.2}$ \\
HD 97334          & HD 97334 B          & $51.5^{+1.7}_{-1.8}$              & $1970\pm20$                 & HIRES & $5890^{+20}_{-32}$      & $4.53^{+0.01}_{-0.01}$ & $+0.20^{+0.02}_{-0.03}$  & $7.73^{+0.4}_{-0.6}$ \\
HII 1348          & HII 1348 B          & $59\pm6$                         & $181^{+49}_{-32}$           & HIRES & $4635^{+36}_{-14}$      & $4.73^{+0.23}_{-0.05}$ & $+0.33^{+0.06}_{-0.01}$ & $13.90^{+0.5}_{-0.4}$ \\
HR 7672           & HR 7672 B           & $75.4\pm0.9$                     & $19^{+1}_{-3}$              & HIRES & $5898^{+1}_{-1}$        & $4.69^{+0.01}_{-0.01}$ & $+0.01^{+0.01}_{-0.01}$ & $2.95^{+0.02}_{-0.02}$ \\
HD 167665         & HD 167665 B         & $60.3\pm0.7$                     & $\sim5.5$                   & HIRES & $6409^{+21}_{-13}$        & $4.24^{+0.05}_{-0.05}$ & $+0.09^{+0.02}_{-0.01}$ & $9.0^{+1.0}_{-1.2}$ \\
HD 33632          & HD 33632 Ab         & $51.7^{+2.6}_{-2.5}$             & $24.1^{+2.0}_{-2.8}$        & HIRES & $6189^{+15}_{-21}$      & $4.39^{+0.05}_{-0.02}$ & $-0.30^{+0.03}_{-0.04}$ & $25^{+4}_{-3}$ \\
HD 49197          & HD 49197 B          & $63^{+13}_{-26}$           & $29.1^{+6.7}_{-9.7}$        & HIRES & $6019^{+27}_{-13}$       & $4.09^{+0.02}_{-0.04}$ & $-0.11^{+0.01}_{-0.01}$ & $40^{+2}_{-3}$ \\
HD 72946          & HD 72946 B          & $69.5\pm0.5$                     & $6.45^{+0.08}_{-0.07}$      & HIRES & $5677^{+12}_{-29}$       & $4.52^{+0.02}_{-0.03}$ & $+0.20^{+0.02}_{-0.02}$ & $2.00^{+0.30}_{-0.50}$ \\
HD 13724          & HD 13724 B          & $50.5\pm3.3$                     & $26.3^{+2.6}_{-0.9}$        & HARPS & $5871^{+31}_{-22}$      & $4.72^{+0.04}_{-0.02}$ & $+0.19^{+0.03}_{-0.02}$ & $3.00^{+0.30}_{-0.5}$ \\
HD 176535         & HD 176535 B         & $65.9^{+2.0}_{-1.7}$             & $\sim13$                    & HARPS & $4698^{+24}_{-33}$      & $4.66^{+0.03}_{-0.02}$ & $-0.03^{+0.01}_{-0.04}$ & $2.20^{+0.20}_{-0.50}$ \\
HD 4113           & HD 4113 C\tablefootmark{c}     & $36\pm5$                         & $50.4^{+2.1}_{-1.4}$        & HARPS & $5701^{+21}_{-13}$      & $4.65^{+0.03}_{-0.04}$ & $+0.14^{+0.03}_{-0.02}$ & $1.54^{+0.22}_{-0.25}$ \\
HD 4747           & HD 4747 B           & $70.0\pm1.6$                     & $10.01\pm0.21$              & HIRES & $5320^{+34}_{-13}$       & $4.76^{+0.03}_{-0.02}$ & $-0.20^{+0.04}_{-0.03}$ & $3.59^{+0.08}_{-0.06}$ \\
HIP 21152         & HIP 21152 B         & $24^{+6}_{-4}$                   & $\sim16$                    & HARPS & $6773^{+31}_{-23}$      & $4.35^{+0.03}_{-0.04}$ & $+0.17^{+0.03}_{-0.01}$ & $46.20^{+2}_{-5}$ \\
AB Pic\tablefootmark{d}            & AB Pic B            & $13.5\pm0.5$         & $242^{+110}_{-56}$          & HARPS & $5285^{+10}_{-9}$       & $4.53^{+0.01}_{-0.01}$ & $+0.04\pm0.02$           & $10.35^{+0.06}_{-0.04}$ \\
HN Peg\tablefootmark{d}            & HN Peg B            & $22.0\pm9.4$       & $795\pm15$                  & HARPS & $6186^{+14}_{-7}$       & $4.48^{+0.03}_{-0.02}$ & $0.00^{+0.01}_{-0.02}$  & $8.73^{+0.06}_{-0.05}$ \\
HR 2562\tablefootmark{d}           & HR 2562 B           & $30\pm15$            & $22.2^{+3.8}_{-2.9}$        & UVES  & $6785^{+29}_{-27}$      & $4.40^{+0.04}_{-0.05}$ & $+0.21^{+0.02}_{-0.03}$  & $43.51^{+0.15}_{-0.17}$ \\
HD 203030\tablefootmark{d}         & HD 203030 B         & $24.1^{+8.4}_{-12}$ & 487                     & HIRES & $5603^{+10}_{-8}$       & $4.64^{+0.03}_{-0.01}$ & $+0.30^{+0.02}_{-0.01}$  & $5.62^{+0.13}_{-0.14}$ \\
CT Cha\tablefootmark{d}            & CT Cha B            & $17\pm6$             & $500^{+320}_{-150}$         & HIRES & $4403^{+6}_{-10}$       & $3.66^{+0.01}_{-0.01}$ & $-0.56^{+0.01}_{-0.01}$ & $13.97^{+0.10}_{-0.15}$ \\
GQ Lup\tablefootmark{d}            & GQ Lup B            & 20                   & $93^{+15}_{-13}$            & HARPS & $4416^{+3}_{-5}$        & $3.65^{+0.01}_{-0.04}$ & $-0.35^{+0.01}_{-0.01}$ & $6.33^{+0.03}_{-0.07}$ \\
ROXs 12\tablefootmark{d}           & ROXs 12 B           & $16\pm4$             & $210\pm20$                  & HIRES & $4059^{+3}_{-4}$        & $3.71^{+0.01}_{-0.01}$ & $+0.14^{+0.01}_{-0.01}$  & $7.20^{+0.03}_{-0.04}$ \\
GSC 06214-00210\tablefootmark{d}   & GSC 06214-00210 B   & $16\pm1$             & 320                         & FEROS & $4119^{+6}_{-13}$       & $3.70^{+0.01}_{-0.04}$ & $-0.06^{+0.01}_{-0.01}$ & $4.24^{+0.04}_{-0.05}$ \\
AF Lep\tablefootmark{e}            & AF Lep b            & $4.3^{+2.9}_{-1.2}$      & $8.98^{+0.15}_{-0.08}$  & HARPS & $6363^{+24}_{-35}$      & $4.58^{+0.04}_{-0.03}$ & $+0.12^{+0.03}_{-0.02}$ & $52.6^{+2.8}_{-3.2}$\\
HD 143811\tablefootmark{e}         & HD 143811 b         & $6.0^{+0.7}_{-0.9}$        & $63.88$                 & HARPS & $6592^{+20}_{-34}$      & $4.80^{+0.04}_{-0.05}$ & $-0.04^{+0.03}_{-0.04}$ & $7.2\pm1.0$ \\
\hline
\end{tabular}
\tablefoot{Separations $s$ are projected separations, except for the systems with a resolved orbit, for which the fitted semimajor axis $a$ is quoted.
\tablefoottext{a}{HIP~74865 is a rapidly rotating early-type star ($v\sin i \gtrsim 75$~km\,s$^{-1}$); the spectroscopic parameters carry larger uncertainties owing to rotational broadening of the absorption lines. The projected rotational velocity is reported as a lower limit since the posterior is pinned at the upper boundary ($80$~km\,s$^{-1}$) of the model grid.}
\tablefoottext{b}{GJ~229~B has been resolved into a close binary (Ba + Bb) with individual masses of $38.1\pm1.0$ and $34.4\pm1.5$~$M_{\rm Jup}$ \citep{Xuan2024Natur}; the listed mass is the total system mass and the semimajor axis refers to the outer orbit around the primary.}
\tablefoottext{c}{The DIBD companion in the HD~4113 system, designated HD~4113~C in the discovery paper \citep{Cheetham2018}.}
\tablefoottext{d}{Atmospheric parameters are adopted from the homogeneous analysis of \citet{2021AJ....161..114S} and reproduced here for completeness; these systems were not refit in the present work.}
\tablefoottext{e}{The directly imaged planetary systems AF~Lep and HD~143811 were discovered after 2020 and were not analyzed by \citet{2021AJ....161..114S}; they are included here as comparison objects.}
}
\end{table*}

\begin{figure*}[ht]
\centering
\includegraphics[width=0.85\textwidth]{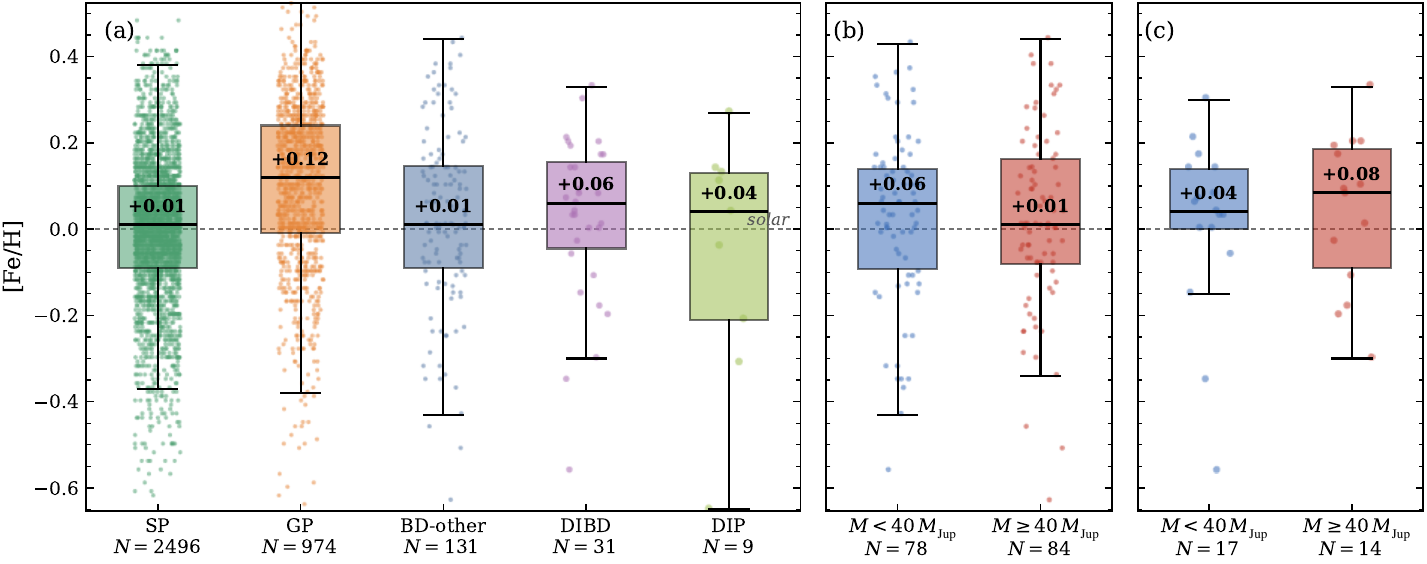}
\caption{Box-and-jitter comparison of host-star [Fe/H] distributions.
\textit{Panel (a)}: Five companion populations: SP hosts ($N=2496$), giant-planet hosts ($N=974$), RV and astrometric
brown-dwarf hosts (BD-other; $N=131$), DIBD
hosts ($N=31$), and hosts of directly imaged planetary-mass companions (DIP; $N=9$). Boxes show the interquartile range, horizontal
lines mark the medians (annotated above each box), and whiskers extend
to $1.5$ times the interquartile range; individual hosts are overplotted as jittered
points, and the dashed gray line marks the solar value.
\textit{Panels (b) and (c)}: BD-other and DIBD samples combined and the DIBD
sample only, each split at a companion mass of $40\,M_{\rm Jup}$.}
\label{fig:boxplot_feh}
\end{figure*}

\begin{figure}[ht]
\centering
\includegraphics[width=\columnwidth]{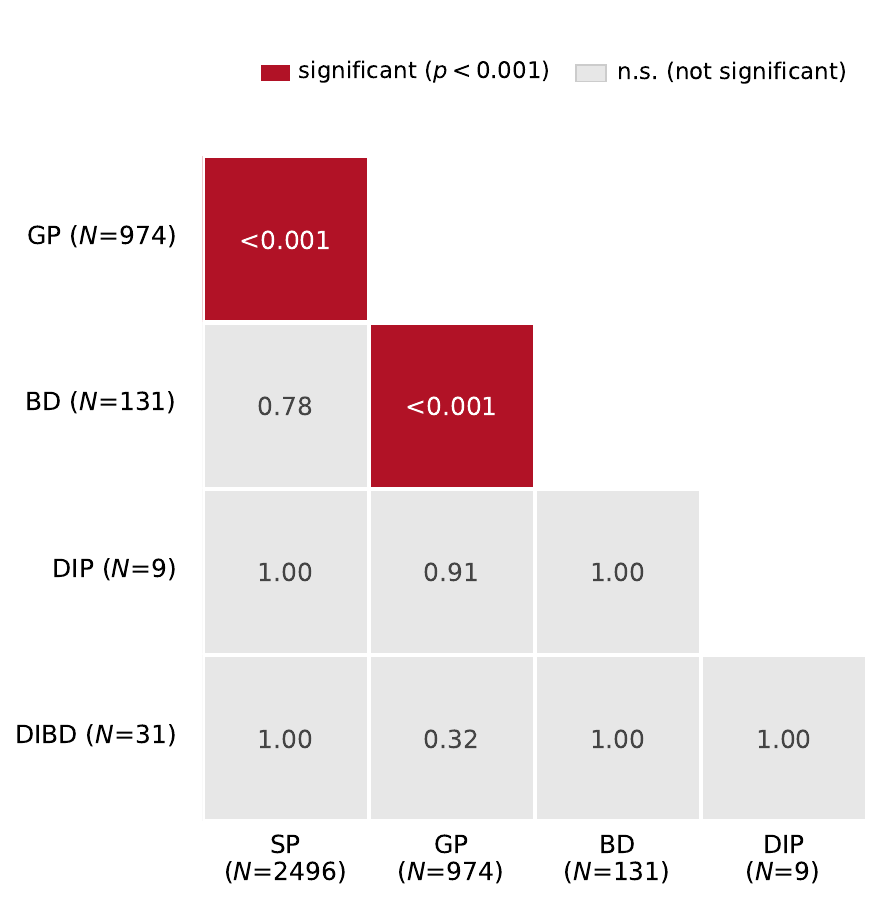}
\caption{Pairwise Mann--Whitney $U$ test on host-star [Fe/H] across
the five populations (``BD'' refers to the close-in BD-other sample,
$N=131$, as in Figs.~\ref{fig:boxplot_feh} and
\ref{fig:cumulative_feh}). Cells show Bonferroni-corrected $p$-values
for the ten pairwise comparisons, color-coded  by significance
(see the legend). Only GP--SP and GP--BD-other survive correction at
$p<0.001$. The label ``n.s.'' denotes a comparison that is not statistically significant after Bonferroni correction.}
\label{fig:mwu_heatmap}
\end{figure}

\subsection{Spectroscopic and kinematic data}
\label{sec:data_spectro_kin}
Of the 54 unique host stars in our DIBD sample, 31 have archival high-resolution
optical spectra from the ESO Science Archive Facility\footnote{\url{http://archive.eso.org}}
or the Keck Observatory Archive\footnote{\url{https://www2.keck.hawaii.edu/koa/public/koa.php}}
from which reliable atmospheric parameters could be derived: 15 observed with
HARPS ($R \approx 115\,000$), 1 with UVES ($R \approx 40\,000$--$110\,000$), and
1 with FEROS ($R \approx 48\,000$) at ESO, and 14 with HIRES ($R \approx 67\,000$)
at Keck. For these 31 systems we derived homogeneous stellar atmospheric
parameters, namely the effective temperature ($T_{\rm eff}$), surface gravity
($\log g$), metallicity ([Fe/H]), and projected rotational velocity ($v\sin i$),
through the Bayesian spectral-synthesis procedure described in
Sect.~\ref{sec:methods_spectro}. The remaining 23 hosts could not be
characterized in this way: either no suitable archival spectrum was available or the available spectra are of predominantly hot and/or rapidly rotating stars
whose spectra contain too few unblended metal absorption lines owing to their
high effective temperatures and/or rotationally broadened line profiles to
constrain the atmospheric parameters reliably. The derived parameters for the 31
systems with atmospheric characterization are listed in
Table~\ref{tab:star_params_all}.
We cross-matched the full sample against the \emph{Gaia}~DR3 catalog
\citep{GaiaDR3_RVs_2023} by first retrieving the cataloged
\emph{Gaia}~DR3 source identifier for each host star from SIMBAD
\citep{Wenger2000}, and then querying the \emph{Gaia} archive on these
identifiers to extract the astrometric and RV parameters.
This procedure yielded a \emph{Gaia}~DR3 counterpart for all 54 host
stars in the sample.

Reliable parallaxes were recovered for 53 of the 54 systems. One
target (Gliese~337) lacks a usable \emph{Gaia}~DR3 parallax and was
therefore excluded from the subsequent kinematic analysis.
\emph{Gaia}~DR3 RVs are available for 47 of the 54 host
stars; the remaining seven systems (HIP~64892, HIP~79098, HIP~81208,
GQ~Lup, $\kappa$~And, HD~151985, and CT~Cha) lack a \emph{Gaia} RV; they are retained for the astrometric (parallax and
proper-motion) analysis but cannot enter the space-velocity
computation. Combining these two cuts, full
three-dimensional Galactic space-velocity components ($U$, $V$, $W$)
are available for 46 of the 54 host stars, following the formalism
described in Sect.~\ref{sec:methods_kin}. The adopted parallaxes,
proper motions, and RVs are reported in
Table~\ref{tab:gaia_params_gt13mj}.

\subsection{Comparison samples}
\label{sec:comparison}

To place the DIBD hosts in context, we compared them with three complementary populations. For planet hosts we adopted the SWEET-Cat catalog \citep{Sousa2021}, which provides homogeneous stellar parameters from consistent spectroscopic analyses; we divided this sample into giant-planet hosts ($M_{\rm p} \geq 0.3~M_{\rm Jup}$) and small-planet (SP) hosts ($M_{\rm p} < 0.3~M_{\rm Jup}$), enabling separate comparisons with the metal-rich population associated with gas-giant formation and the low-mass regime where core accretion operates under different conditions. For close-in brown-dwarf companions we used the compilation of \citet{Stevenson2023}, who listed 214 confirmed brown-dwarf companions with orbital periods $P < 10^4$~days detected primarily through RV and transit surveys; counting each host only once, excluding hosts without a reported metallicity measurement, and restricting to companions in the $13$--$80~M_{\rm Jup}$ range yields a comparison sample of $N=131$ host stars (hereafter ``BD-other''), which probe the short-period regime of the desert.

In addition, we included directly imaged planetary-mass companions from our earlier analysis \citep{2021AJ....161..114S} as well as newly discovered systems not covered in that work (Table~\ref{tab:star_params_all}). This combined dataset allows us to investigate whether the host-star properties of DIBDs differ systematically from those of GP hosts, SP hosts, close-in brown-dwarf systems, and DIP hosts.

\begin{figure*}[t]
\sidecaption
\parbox[b]{12cm}{\includegraphics[width=12cm]{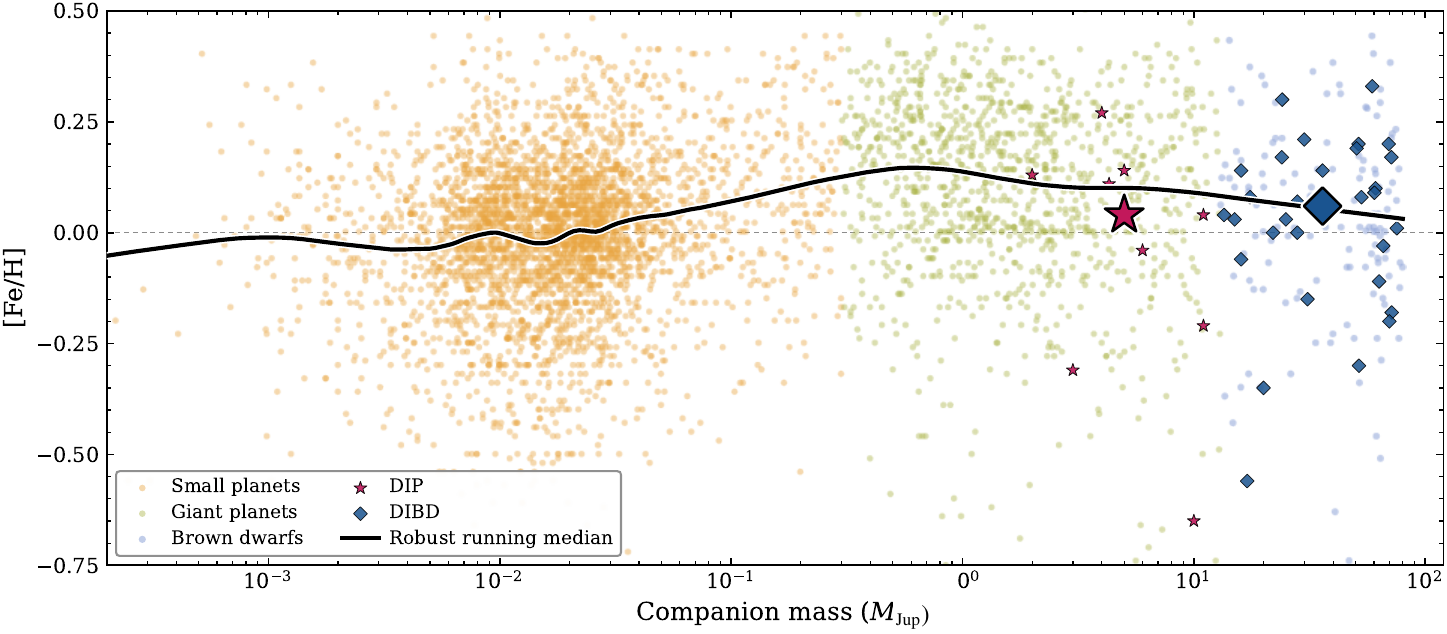}\\[2pt]\includegraphics[width=12cm]{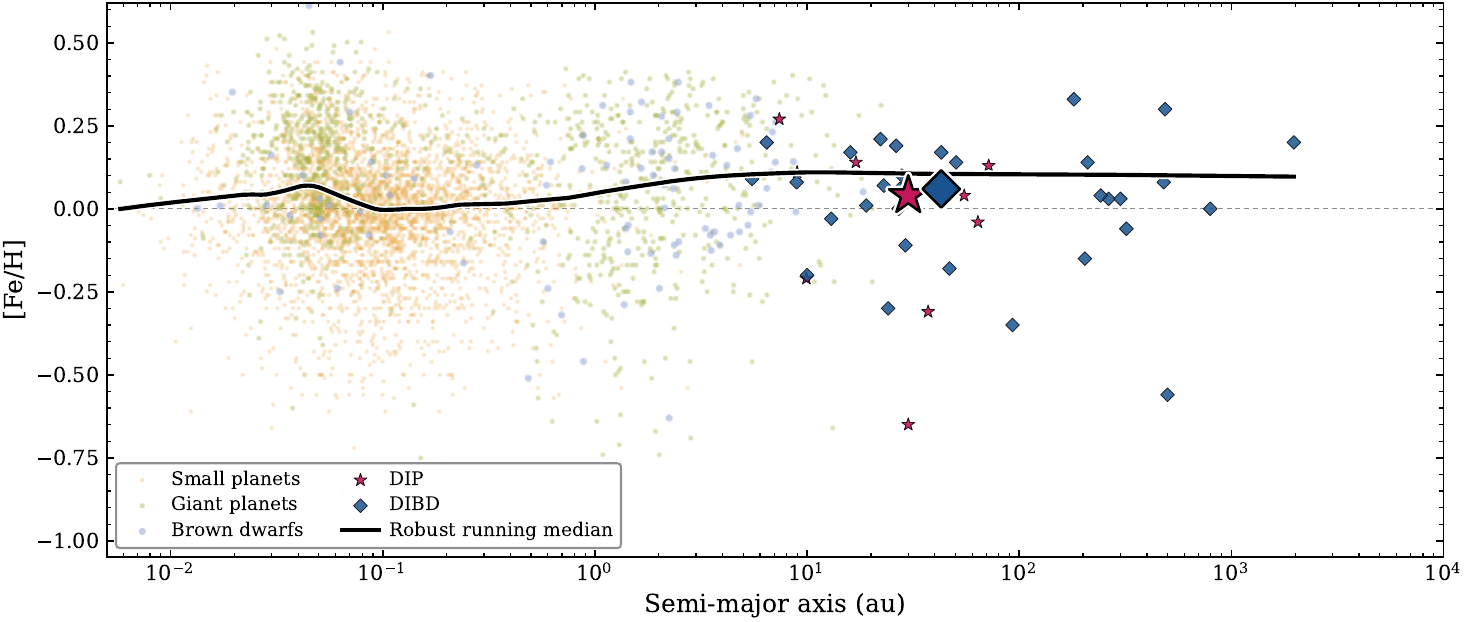}}
\caption{Host-star metallicity for SPs (orange), GPs (green), brown dwarfs from RV and astrometric surveys (light
blue), DIP companions (red stars), and DIBD companions (dark blue
diamonds), shown as a function of companion mass (\textit{top}) and
projected separation, $s$ (\textit{bottom}). Large filled symbols mark
the DIP and DIBD sample medians, and the solid black curve is a robust
running median across the combined SP+GP+BD distribution. The DIP and
DIBD medians sit essentially on the running-median curve, which peaks
near $1\,M_{\rm Jup}$ in mass and is flat in separation; the directly
imaged populations populate the wide-separation tail
($s \gtrsim 5$\,au) and span the full range of host metallicities seen
at smaller separations.}
\label{fig:met_mass}
\end{figure*}
\subsection{Spectroscopic analysis}
\label{sec:methods_spectro}
For the subset of 31 host stars with archival high-resolution optical spectra
(Sect.~\ref{sec:data_spectro_kin}), we derived the stellar atmospheric
parameters, namely the effective temperature ($T_{\rm eff}$), surface gravity
($\log g$), metallicity ([Fe/H]), and projected rotational velocity ($v\sin i$),
through spectral synthesis. The instruments used and the parameters derived for
each system are listed in Table~\ref{tab:star_params_all}.

The analysis was carried out with the \texttt{iSpec} framework
\citep{Blanco2014a,Blanco2014b}, following the methodology of
\citet{2021AJ....161..114S}. Synthetic spectra were computed from MARCS model
atmospheres \citep{Gustafsson2008} with the radiative-transfer code
\texttt{SPECTRUM} \citep{Gray1994} and the atomic line list distributed with
\texttt{iSpec}. We used the same diagnostic wavelength intervals as \citet{2021AJ....161..114S}: the Mg\,\textsc{i} triplet region at $515$--$520$~nm, which is sensitive to $\log g$; the $600$--$620$~nm region, which contains numerous relatively isolated lines of Fe\,\textsc{i}, Fe\,\textsc{ii}, Ti\,\textsc{i}, Ni\,\textsc{i}, and Ca\,\textsc{i} and constrains [Fe/H] and $v\sin i$; and the H$\alpha$ region at $654$--$659$~nm, whose wings constrain $T_{\rm eff}$. For spectra affected by H$\alpha$ emission or veiling, we used the emission-free $590$--$596.5$~nm interval and, where appropriate, the $610$--$620$~nm interval. The observed spectra were fitted iteratively with these synthetic
spectra, simultaneously varying $T_{\rm eff}$, $\log g$, [Fe/H], and the
projected rotational velocity $v\sin i$. The parameter space is sampled with the
Markov chain Monte Carlo (MCMC) ensemble sampler \texttt{emcee}
\citep{Foreman-Mackey2013}, which yields the full posterior probability
distributions of the parameters together with their covariances; we adopted the
posterior medians as the best-fit values.

This procedure provides robust parameters for the F-, G-, and K-type stars that
make up the majority of the spectroscopic subsample. The few early-type hosts
are more demanding. For the A-type star HIP~74865 ($v\sin i \gtrsim
75$~km~s$^{-1}$), for example, rapid rotation blends most lines and leaves few
features suitable for synthesis. For such stars the same procedure was applied,
with [Fe/H] constrained from the subset of comparatively unblended metal lines
in the spectral regions least affected by rotational broadening.

For the FGK-type stars, the typical internal uncertainties from the MCMC
posteriors are $\sigma(T_{\rm eff}) \approx 50$~K, $\sigma(\log g) \approx
0.05$~dex, and $\sigma({\rm [Fe/H]}) \approx 0.03$~dex, consistent with the
precision achieved in our previous analysis of DIP hosts
\citep{2021AJ....161..114S}. We caution that these internal MCMC uncertainties
are statistical lower limits; the realistic error budget is dominated by
systematic effects (the adopted model atmospheres, atomic data, and analysis
method). Benchmark comparisons of synthesis-based atmospheric parameters
indicate typical systematic uncertainties of order $\sim50$--$100$~K in
$T_{\rm eff}$, $\sim0.1$~dex in $\log g$, and $\sim0.05$--$0.1$~dex in [Fe/H]
\citep{Blanco2014a,Jofre2014,Heiter2015}, which we adopted as conservative
external error estimates.

\subsection{Kinematic analysis}
\label{sec:methods_kin}

Galactic space-velocity components characterize stellar population membership (thin disk, thick disk, or halo) and provide a consistency check on the strong youth bias of the directly imaged sample. We computed the velocity components $U$, $V$, $W$ for each DIBD host star using astrometric and spectroscopic data from \emph{Gaia}~DR3 \citep{GaiaDR3_DiscKinematics_2023}: parallaxes, proper motions ($\mu_\alpha\cos\delta$, $\mu_\delta$), and RVs were taken directly from the \emph{Gaia} catalog. We adopted the standard right-handed convention in which $U$ is positive toward the Galactic center, $V$ in the direction of Galactic rotation, and $W$ toward the north Galactic pole, and corrected to the local standard of rest (LSR) using the solar motion $(U, V, W)_\odot = (11.1,\, 12.2,\, 7.3)$~km~s$^{-1}$ \citep{Schoenrich2010}.

For each star we computed the peculiar velocity relative to the LSR,
\begin{equation}
v_{\rm pec} = \sqrt{U_{\rm LSR}^2 + V_{\rm LSR}^2 + W_{\rm LSR}^2}\,,
\end{equation}
which serves as a proxy for population membership: thin-disk stars typically have $v_{\rm pec} \lesssim 40$~km~s$^{-1}$, thick-disk stars reach $v_{\rm pec} \sim 50$--$100$~km~s$^{-1}$, and halo stars exhibit still higher velocities. Velocity dispersion is also correlated with age through dynamical heating of the Galactic disk \citep[e.g.,][]{Casagrande2011,2022AJ....164..181U,2025AJ....169...13P}. 
To provide context for the DIBD kinematics, we computed $U$, $V$, $W$ for the comparison samples using the same procedure. For GP hosts, we drew on existing tabulations \citep[e.g.,][]{Ghezzi2010,2023AJ....166...91S} supplemented with our own calculations from \emph{Gaia}~DR3. For the 214 RV and transit brown-dwarf systems from \citet{Stevenson2023}, we cross-matched the retained host-star sample with \emph{Gaia}~DR3 and derived velocities in the same manner, excluding systems with missing or unreliable data. We visualized the kinematic distributions using the Toomre diagram (Fig.~\ref{toomre}), which plots $\sqrt{U_{\rm LSR}^2 + W_{\rm LSR}^2}$ versus $V_{\rm LSR}$ and provides a classical diagnostic for separating thin-disk, thick-disk, and halo populations \citep[e.g.,][]{2003A&A...410..527B}. We also computed the median peculiar velocity and three-dimensional velocity dispersion for each population subsample to enable quantitative comparisons. For stars in common with our previous kinematic analysis of exoplanet host stars \citep{2023AJ....166...91S}, the recovered $U,V,W$ velocities agree to within $\sim 1$--$2$~km\,s$^{-1}$, with residual differences traceable to small variations in the adopted solar motion and \emph{Gaia} source-matching conventions, confirming the consistency of our pipeline. The velocities confirm that the imaged hosts are predominantly young, thin-disk stars, as expected from the selection function of direct imaging; these results are discussed in Sect.~\ref{sec:results_kinematics}.

\section{Results}
\label{sec:results}

\subsection{Atmospheric parameters of the DIBD host sample}
\label{sec:results_params}

Table~\ref{tab:star_params_all} summarizes the atmospheric parameters for the 31 DIBD hosts with suitable high-resolution spectra. Eight are from \citet{2021AJ....161..114S}, while the remaining 23 are newly analyzed in this work. Two newly analyzed planetary-mass comparison systems are included at the end of the table. The sample covers a wide range of stellar properties: effective temperatures
span $T_{\rm eff} \approx 3800$--$6800$~K, corresponding to early-M through
mid-F spectral types, while surface gravities lie in the range $\log g \approx
3.6$--$5.0$~dex (cgs), encompassing both main-sequence dwarfs and the younger,
less compact pre-main-sequence members of nearby moving groups (CT\,Cha,
GQ\,Lup, ROXs\,12, and GSC\,06214$-$00210). Projected rotational velocities span
more than an order of magnitude, from $v\sin i \lesssim 5$~km\,s$^{-1}$ for the
slow-rotating G-, K-, and M-type dwarfs (HD\,13724, GJ\,758, GJ\,229, and
HD\,72946) to $v\sin i \approx 70$--$80$~km\,s$^{-1}$ for the fast-rotating
early-type hosts (HIP\,74865 and PZ\,Tel). Host-star metallicities range from
[Fe/H]~$=-0.56$~dex for the metal-poor T\,Tauri host CT\,Cha to
[Fe/H]~$=+0.33$~dex for the metal-rich Pleiades host HII\,1348, with the bulk of
the sample concentrated within $\pm 0.2$~dex of the solar value.

\subsection{The host-star metallicity distribution}
\label{sec:results_metallicity}

The metallicity distribution of the DIBD host sample is shown in
Fig.~\ref{fig:feh_hist}. The distribution is broad and approximately
symmetric, peaking near solar metallicity, with a median of
[Fe/H]~$=+0.06$~dex and a robust median absolute deviation of $0.11$~dex
(mean $+0.02$~dex, standard deviation $0.19$~dex); three of the 31 hosts (CT\,Cha, GQ\,Lup, and HD\,33632) lie
at or below [Fe/H]~$=-0.3$. There is no indication of bimodality or of an
excess of metal-rich hosts in the sample.

To place the DIBD sample in the broader context of substellar
companion populations, we compared its metallicity distribution to
those of four reference samples: SP hosts ($M_{\rm p}<0.3\,M_{\rm Jup}$;
SP, $N=2496$), GP hosts ($0.3$--$13\,M_{\rm Jup}$;
GP, $N=974$), brown-dwarf companions detected primarily through
RV and astrometric techniques (BD-other, $N=131$), and directly imaged
planetary-mass companions (DIP, $N=9$). The reference samples include
one entry per host star and are described in Sect.~\ref{sec:comparison}. For
each pair of populations we tested the null hypothesis that the two
[Fe/H] distributions are drawn from the same parent population
using a two-sided Mann--Whitney $U$  test \citep{MannWhitney1947}, a
nonparametric rank-based test of whether one of two samples is
stochastically larger than the other; unlike the two-sample $t$-test
it does not assume Gaussian distributions and is therefore robust
to the heavy tails and outliers seen in metallicity samples.
Significance was assessed at $p<0.001$ after Bonferroni correction
\citep{Bonferroni1936,Dunn1961} for the ten pairwise comparisons. The side-by-side comparison is shown as a five-panel
box-and-jitter plot in Fig.~\ref{fig:boxplot_feh}a, and the full matrix of
pairwise $p$-values is given in Fig.~\ref{fig:mwu_heatmap}. Only two of the ten
comparisons are significant after correction: GP hosts are more
metal-rich than both SP hosts (corrected $p\sim10^{-52}$) and BD-other
hosts (corrected $p\approx6\times10^{-4}$); all eight remaining comparisons,
including every comparison involving the DIBD sample, do not show a statistically significant difference at the present sample sizes. We caution that the reference samples are highly unbalanced in size (from $N=2496$ for SPs to $N=9$ for DIPs), so the Mann--Whitney comparisons involving the smallest samples (DIP, and to a lesser extent DIBD) have limited statistical power and should therefore be interpreted as exploratory. For these samples, a nonsignificant result may reflect limited statistical power rather than intrinsic similarity. Moreover, whereas the metallicities derived in this work (DIBD, DIP, and the SWEET-Cat planet hosts) are homogeneous, the BD-other metallicities are compiled from heterogeneous literature sources \citep{Stevenson2023}; differences involving that sample, or their absence, may partly reflect systematic offsets between analyses rather than intrinsic population differences.

The GP sample stands apart
from the SP and BD-other samples: its [Fe/H] distribution is shifted as a
whole toward higher metallicities, with a median of $+0.12$~dex and an
interquartile range that lies almost entirely above the solar value.
The remaining four populations, SP, BD-other, DIBD, and DIP, have
medians of $+0.01$, $+0.01$, $+0.06$, and $+0.04$~dex, respectively,
all within $\pm 0.06$~dex of solar, and overlapping interquartile
ranges that straddle the solar value. The DIBD interquartile range
extends from approximately $-0.05$ to $+0.16$~dex, comparable to that
of the SP and GP samples and slightly narrower
than the BD-other sample, reflecting the diversity of stellar masses
and formation environments represented in the directly imaged
hosts.

\subsection{Dependence on companion mass}
\label{sec:results_mass}

We next investigated whether the host-star metallicity distribution
varies with companion mass within the brown-dwarf regime. Following
the mass scale adopted in earlier studies of the brown-dwarf desert
\citep{Ma2014,Maldonado2017}, we partitioned the BD-other and DIBD
samples at $M_{\rm comp}=40\,M_{\rm Jup}$, a value that approximately
divides the bottom of the brown-dwarf mass function from the
high-mass tail dominated by deuterium-burning, more massive
substellar companions. The corresponding two-bin comparison is shown
in Figs.~\ref{fig:boxplot_feh}b and \ref{fig:boxplot_feh}c.

Figure~\ref{fig:boxplot_feh}b shows the result for the combined BD-other$+$DIBD sample
(i.e.,\ close-in and DIBD hosts taken together). Of the 162 hosts in this combined sample, all have a tabulated companion mass and enter the mass split.
The two subgroups have nearly identical metallicity distributions:
companions below $40\,M_{\rm Jup}$ ($N=78$) have a median host
metallicity of [Fe/H]~$=+0.06$, while those at or above
$40\,M_{\rm Jup}$ ($N=84$) have a median of $+0.01$, with strongly
overlapping interquartile ranges. The same exercise restricted to the
DIBD sample alone (panel c) yields medians of $+0.04$ for the 17
hosts of lower-mass companions and $+0.08$ for the 14 hosts of
higher-mass companions. In both panels the medians differ by no
more than $0.05$~dex, and the boxes overlap almost entirely. We do not detect a metallicity dependence on companion mass within the
brown-dwarf regime in either the larger RV and astrometric sample or in
the directly imaged subsample.

A complementary view is provided by Fig.~\ref{fig:met_mass}, which
shows host metallicity as a function of companion mass, from terrestrial-mass planets to
the stellar boundary.
The solid black curve traces a robust running median across the
combined SP$+$GP$+$BD distribution, computed with Cleveland's
robust local regression \citep[][]{Cleveland1979} in log-mass
space (and in log-separation space for the lower panel). The curve uses the nearest $25\%$ of points at each location and
iteratively down-weights outliers, so that it follows the bulk of the
distribution and is insensitive to a handful of extreme [Fe/H] values. The running median rises
gradually from $\sim 0.00$~dex at $M_{\rm comp} \lesssim
10^{-2}\,M_{\rm Jup}$ to a maximum of $+0.12$~dex near
$M_{\rm comp} \approx 1\,M_{\rm Jup}$, where the GP
metallicity excess dominates, and then declines through the
intermediate brown-dwarf regime back toward solar values, reaching
[Fe/H]~$\approx +0.03$ by $M_{\rm comp} \approx 80\,M_{\rm Jup}$, the upper
edge of the brown-dwarf range. The
DIBD points (dark blue diamonds) and DIP points (red stars) are
broadly distributed in [Fe/H] at their respective masses, and their
sample medians (large filled symbols) sit within $\sim 0.05$~dex of
the running median curve. The directly imaged populations therefore
follow rather than depart from the overall mass--metallicity trend
defined by the much larger close-in samples.

\subsection{Dependence on projected orbital separation}
\label{sec:results_sma}

The directly imaged populations probe a region of orbital
parameter space that is largely inaccessible to RV and transit
surveys. The bottom panel of Fig.~\ref{fig:met_mass} shows host metallicity as a function
of projected separation for all populations. The DIBD and DIP
companions populate the wide-separation tail of the distribution, with
$s$ ranging from $\sim 5$~au to $\sim 2000$~au, and exhibit the full
range of host metallicities seen at smaller separations:
$-0.56 \leq [\mathrm{Fe/H}] \leq +0.33$ for the DIBD sample and
$-0.65 \leq [\mathrm{Fe/H}] \leq +0.27$ for the DIP sample. The
robust running median is essentially flat at
[Fe/H]~$\approx +0.03$--$+0.09$~dex across the wide-separation regime
($s \gtrsim 1$~au), and we identify no monotonic trend of host
metallicity with projected separation either within the directly imaged
sample taken in isolation or across the full combined sample.

\begin{figure*}[t]
\sidecaption
\includegraphics[width=10cm]{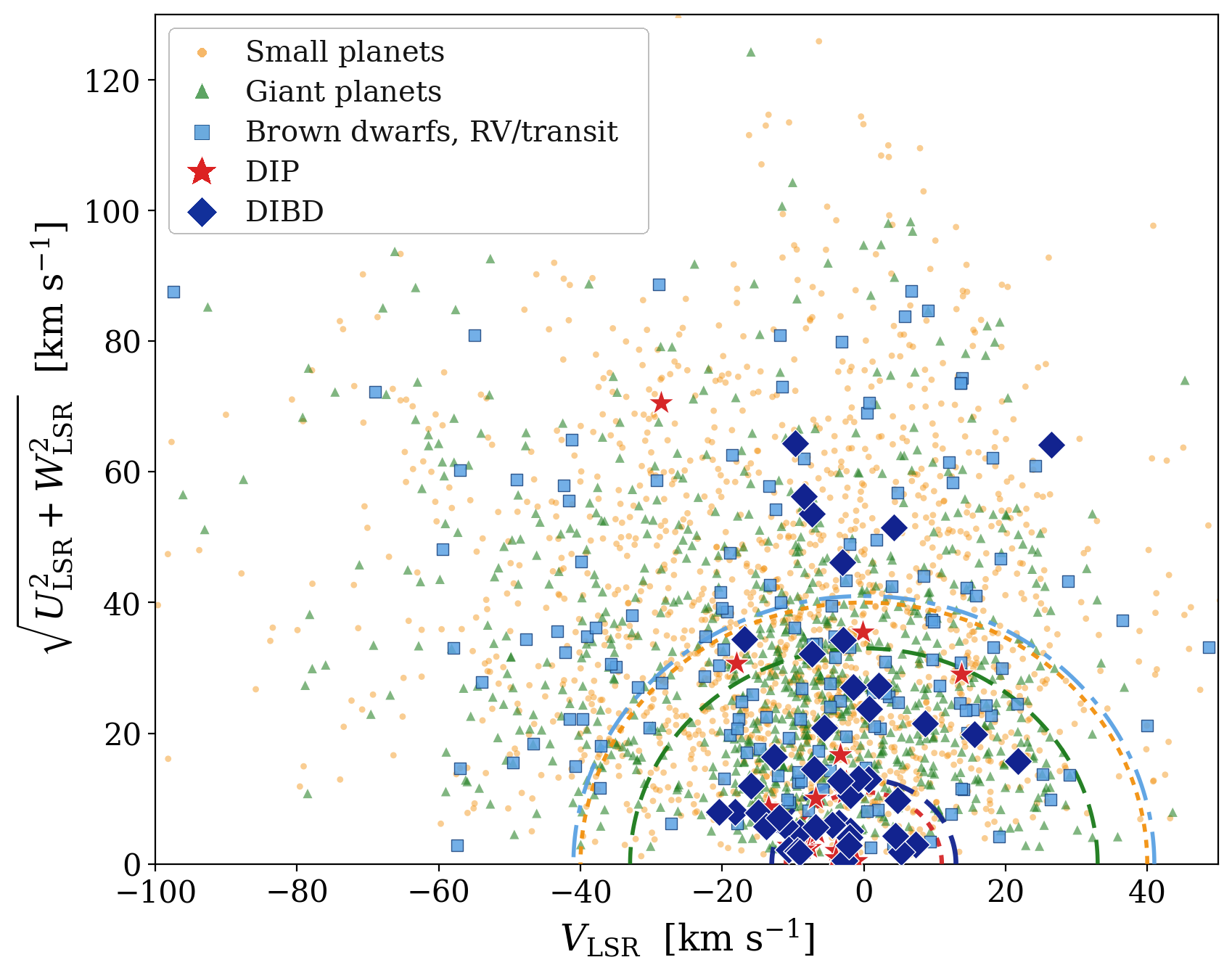}
\caption{Toomre diagram for the hosts of planetary and substellar companions. The dashed semicircles are curves of constant peculiar velocity drawn at the median $v_{\rm pec}$ of each sample (Table~\ref{tab:kinematics}), color-coded as in the legend; they are not population boundaries.}
\label{toomre}
\end{figure*}

\begin{table}[t]
\caption{Kinematic properties of the companion-host samples.}
\label{tab:kinematics}
\centering
\small
\setlength{\tabcolsep}{4pt}
\begin{tabular}{lcc}
\hline\hline
Population & $v_{\rm pec}$ (km s$^{-1}$) & $\sigma_v$ (km s$^{-1}$) \\
\hline
Giant planets (GP) & 33 & 42 \\
Small planets (SP) & 40 & 50 \\
Directly imaged planets (DIP) & 11 & 23 \\
Directly imaged brown dwarfs (DIBD) & 13 & 23 \\
\quad $M_{\rm comp}<40~M_{\rm Jup}$ & 10 & 16 \\
\quad $M_{\rm comp}\geq40~M_{\rm Jup}$ & 22 & 30 \\
BD-other & 41 & 63 \\
\hline
\end{tabular}
\tablefoot{The two indented rows are overlapping mass-defined subsets of the full DIBD sample and are not independent populations.}
\end{table}
\subsection{Galactic kinematics of DIBD host stars}
\label{sec:results_kinematics}

We computed Galactic velocity components for 46 of the 54 unique DIBD host stars. Figure~\ref{toomre} presents the Toomre diagram for the DIBD hosts together with the comparison samples. The DIBD host stars are concentrated in the thin-disk region, clustering at low peculiar velocity, as expected given the large fraction of young, nearby systems in direct-imaging surveys. Their velocity distribution is in fact more compact than that of the GP hosts, and no object in our sample reaches halo-like velocities.

To quantify the differences between the samples, we computed median peculiar velocities and three-dimensional velocity dispersions as listed in Table~\ref{tab:kinematics}. The GP hosts have a median peculiar velocity of $\sim33$~km~s$^{-1}$ and a velocity dispersion of $\sim42$~km~s$^{-1}$. The DIBD host sample is kinematically colder, with a median peculiar velocity of $\sim13$~km~s$^{-1}$ and a dispersion of $\sim23$~km~s$^{-1}$. The RV and transit brown dwarf hosts, by contrast, are kinematically warmer (median $\sim41$~km~s$^{-1}$, dispersion $\sim63$~km~s$^{-1}$), as expected for an older field population.

We also divided the DIBD hosts at $M_{\rm comp}=40~M_{\rm Jup}$, matching the split used in the metallicity analysis (Sect.~\ref{sec:results_mass}). Both subgroups are kinematically cold: hosts of higher-mass brown dwarfs have a median peculiar velocity of about $22$~km~s$^{-1}$ and hosts of lower-mass companions about $10$~km~s$^{-1}$, with dispersions of approximately $30$ and $16$~km~s$^{-1}$, respectively. These estimates rest on small subsamples and should be interpreted with caution, but neither subgroup shows the elevated velocities characteristic of an old disk population.

Overall, the kinematic analysis shows that DIBD host stars are predominantly young, thin-disk objects, kinematically colder than both the GP hosts and the close-in (RV and transit) brown-dwarf hosts. This is consistent with the strong youth bias of direct-imaging selection. The velocity contrast with the close-in brown-dwarf sample is therefore driven by selection (young imaged hosts vs. older field hosts), not by origin, and cannot be used as evidence of a formation channel.

\begin{figure}[ht]
\centering
\includegraphics[width=\columnwidth]{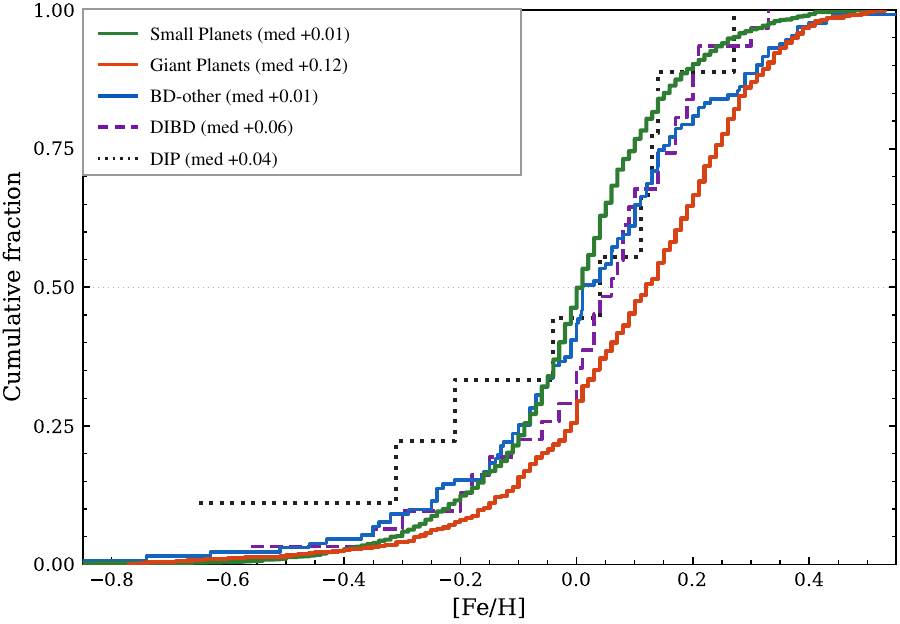}
\caption{Cumulative [Fe/H] distributions of host stars for the five
companion populations. The median metallicity of each sample is given in the
legend. The GP hosts (orange) are systematically offset
toward higher [Fe/H], while the BD-other, DIBD, and DIP curves track
the SP distribution closely. The horizontal dotted line marks the 50\% (median) level.}
\label{fig:cumulative_feh}
\end{figure}

\section{Discussion}
\label{sec:discussion}

\subsection{Formation channels and observational diagnostics}
\label{sec:disc_diagnostics}

Three formation channels can in principle produce brown-dwarf
companions: core accretion in a protoplanetary disk, gravitational
instability of the disk itself, and star-like collapse of the parent
molecular cloud
\citep{Pollack1996,Bate2002,Stamatellos2009,Chabrier2014}. Their
expected parameter-space regimes, in mass ratio $q = M_{\rm
comp}/M_\star$, separation $a$, and host metallicity, are summarized
in Table~\ref{tab:formation_mechanism}. Core accretion is the only
channel with a strong, established metallicity dependence
\citep{Fischer2005,Mordasini2012}, reflecting its sensitivity to the
solid reservoir of the protoplanetary disk: a more metal-rich disk
provides more material for efficient core growth. Disk instability and cloud fragmentation depend
mainly on local disk and cloud thermodynamics and are essentially
metallicity-independent at the population level
\citep{Bate2002,Stamatellos2009,Chabrier2014}, although a weak residual
opacity-driven dependence is expected through the cooling efficiency
of the parent gas. The metallicity
distributions of different companion populations should
carry fingerprints of which channel dominates in each mass and
separation regime.

The representative ranges in Table~\ref{tab:formation_mechanism} are drawn from theoretical models and population studies on core accretion (\citealt{Pollack1996,IdaLin2004,Mordasini2012,Benz2014}), disk instability (\citealt{Toomre1964,Rice2005,Rafikov2005,Stamatellos2009}), and cloud fragmentation (\citealt{Bate2002,Chabrier2014}). These regimes overlap substantially, and post-formation migration or scattering can blur primordial signatures \citep{Chabrier2014}. As
emphasized in Sect.~\ref{sec:comparison} and Appendix~\ref{app:formation},
we treated individual classifications as indicative rather than
definitive; the meaningful unit is the population-level distribution.
We use this framework in the following subsections to interpret the
host metallicity distribution (Sect.~\ref{sec:disc_metallicity_test}),
its dependence on companion mass within the brown-dwarf regime
(Sect.~\ref{sec:disc_mass_split}), and the close-orbit deficit
(Sect.~\ref{sec:disc_desert}).

\begin{table*}[t]
\centering
\caption{Heuristic parameter-space regimes for brown-dwarf companion formation pathways.}
\label{tab:formation_mechanism}
\begin{tabular}{lcc>{\raggedright\arraybackslash}p{0.32\textwidth}}
\hline\hline
Formation pathway & $q$ & $a$ (au) & Host-metallicity expectation \\
\hline
Core accretion & $\lesssim0.01$ & $\lesssim20$--30 & Strong metal-rich preference \\
Disk instability & $\sim0.01$--0.1 & $\sim50$--100 & Weak or absent dependence \\
Cloud fragmentation & $\gtrsim0.02$ & $\gtrsim50$ & No planet-like metallicity~preference \\
\hline
\end{tabular}
\tablefoot{The quoted separations are the model-predicted semimajor axes ($a$) at which each
channel operates; the separations measured in this work are projected separations $s$
(Sect.~\ref{sec:sample_selection}).}
\end{table*}

\subsection{Host metallicity and the mass--metallicity arc}
\label{sec:disc_metallicity_test}

The cumulative metallicity distributions in
Fig.~\ref{fig:cumulative_feh} show the GP hosts
systematically shifted toward higher [Fe/H], reaching the 50\% level
only at [Fe/H]~$\approx +0.12$, while the SP, BD-other,
DIBD, and DIP curves all reach the median between [Fe/H]~$\approx 0$
and $+0.06$. The DIBD and BD-other curves track each other closely from below
$-0.3$~dex to above $+0.3$~dex. Pairwise Mann--Whitney tests with
Bonferroni correction (Fig.~\ref{fig:mwu_heatmap}) confirm this
picture: the only comparisons that pass at $p<0.001$ are SP--GP (the
classical planet--metallicity correlation; \citealt{Gonzalez1997,Fischer2005})
and GP--BD-other. No statistically significant difference is detected for the other pairs
at the current sample sizes; only the GP sample is offset
from the others. The DIBD--GP comparison is informative
but does not survive correction at our sample size: the DIBD median
($+0.06$) is offset from the GP median ($+0.12$) by $\sim0.06$~dex,
smaller than the $\sim0.11$~dex offset that separates BD-other from
GP but qualitatively in the same direction; only the GP sample,
sampled with $N=131$ versus $N=31$, crosses the threshold.

The consistency between DIBD and BD-other, and the joint offset of
both from GP hosts, indicates that the host-metallicity
signature found for close-in RV brown dwarfs
\citep{Ma2014,Maldonado2017} extends to the wide-orbit, directly
imaged regime. The mass--metallicity arc in Fig.~\ref{fig:met_mass}
shows this directly, with the DIBD and DIP medians sitting on the
running-median curve at their respective masses. We note, however, that the near-solar metallicity of brown-dwarf hosts
also overlaps with that of hosts of low-eccentricity cold Jupiters, which
show approximately solar metallicities in recent homogeneous studies
\citep[e.g.,][]{Buchhave2018,Banerjee2024}. This comparison is not
universal across giant-planet architectures, as high-eccentricity cold-Jupiter
hosts can be substantially more metal-rich \citep{Banerjee2024}.
Companion atmospheric abundances provide a complementary formation diagnostic:
HR~7672~B has C and O abundances consistent with those of its primary star
\citep{2022AJ....163..189W}, whereas directly imaged planetary-mass companions
can show superstellar atmospheric metallicities consistent with substantial early
solid accretion \citep{2025ApJ...981..138W}.
Metallicity alone does not exclude a contribution from core accretion in the
lower-mass brown-dwarf regime. Taken together, the available
evidence is consistent with disk instability or cloud fragmentation, but does
not distinguish between these two formation channels.

A natural extension of this comparison is toward higher companion masses. Wide FGK$+$M
binaries provide a homogeneously analyzed sample of primaries spanning the low-mass stellar
regime \citep{Montes2018,DuqueArribas2024}, and merging such samples with the substellar
companions studied here would allow the mass--metallicity comparison to be extended
continuously from the deuterium-burning limit to $\sim0.6\,M_\odot$.

\subsection{No mass dependence within the brown-dwarf regime}
\label{sec:disc_mass_split}

A second question is whether the brown-dwarf population is internally
homogeneous. The $\sim40$--$45\,M_{\rm Jup}$ boundary
\citep{Ma2014,Maldonado2017} is motivated empirically rather than by a
single physical threshold: it coincides with the driest part of the
brown-dwarf desert and with a transition in the companion eccentricity
distribution near $\sim42.5\,M_{\rm Jup}$, from the low eccentricities
typical of disk-formed, planet-like objects below the boundary to the
higher, more stellar-like eccentricities above it \citep{Ma2014}, and
with a weakening of the host-metallicity correlation toward higher
masses \citep{Maldonado2017}. It was thus proposed as the mass above
which star-like fragmentation dominates and below which a tail of
core-accretion products could remain present; both disk instability
and cloud fragmentation are essentially metallicity-independent, so a
metallicity step across this boundary would only be expected if a
core-accretion component persists at the lower-mass end. Adopting a
$40\,M_{\rm Jup}$ split, the lower- and higher-mass DIBD subgroups have
closely similar median metallicities ($+0.04$ vs.\ $+0.08$; Mann--Whitney $U$  test,
$p=0.65$), as do the combined BD-other$+$DIBD subgroups ($+0.06$ vs.\
$+0.01$, $p=0.70$). We do not recover the $\sim 0.2$~dex offset
reported by \citet{Maldonado2017}, which we attribute to (i) recently
characterized low-mass DIBDs at near-solar metallicity, and (ii)
revised dynamical masses that have shifted several systems across the boundary.
The absence of a metallicity step is consistent with a
negligible core-accretion contribution across the brown-dwarf mass
range, and does not by itself distinguish disk instability from cloud
fragmentation.

A null result is not the same as ruling out two formation channels.
With $N=31$, our sensitivity to a $\sim0.1$~dex difference between
subpopulations is limited by the intrinsic $0.19$~dex scatter of
host metallicities. A complication also remains: the lower-mass DIBDs
include a substantial fraction of young ($<100$~Myr) companions
identified by blind imaging of nearby moving groups, drawn from
metallicity-narrow parent populations, while the higher-mass DIBDs
are mostly older field FGK stars selected through long-baseline RV
trends and \textsc{Hipparcos}--\emph{Gaia} astrometric accelerations,
drawn from a more metallicity-diverse pool. Any mass-dependent
formation signature could be partially masked by this
contrast between host populations.

\subsection{The brown-dwarf desert at close orbits}
\label{sec:disc_desert}

The classical brown-dwarf desert at close orbits is well established
in RV and transit surveys at $P \lesssim 100$~days and
$\sim 35$--$55\,M_{\rm Jup}$ \citep{Marcy2000,Grether2006,Ma2014}, and
is recovered in our sample at the same location in the
mass--separation plane (Fig.~\ref{mass_sem}, $s \lesssim 1$~au).
The directly imaged population shows the complementary feature: the
same $\sim 25$--$50\,M_{\rm Jup}$ mass range that is depleted at close
orbits is well populated at projected separations of tens to hundreds
of au.

\begin{figure}
\centering
\includegraphics[width=\columnwidth]{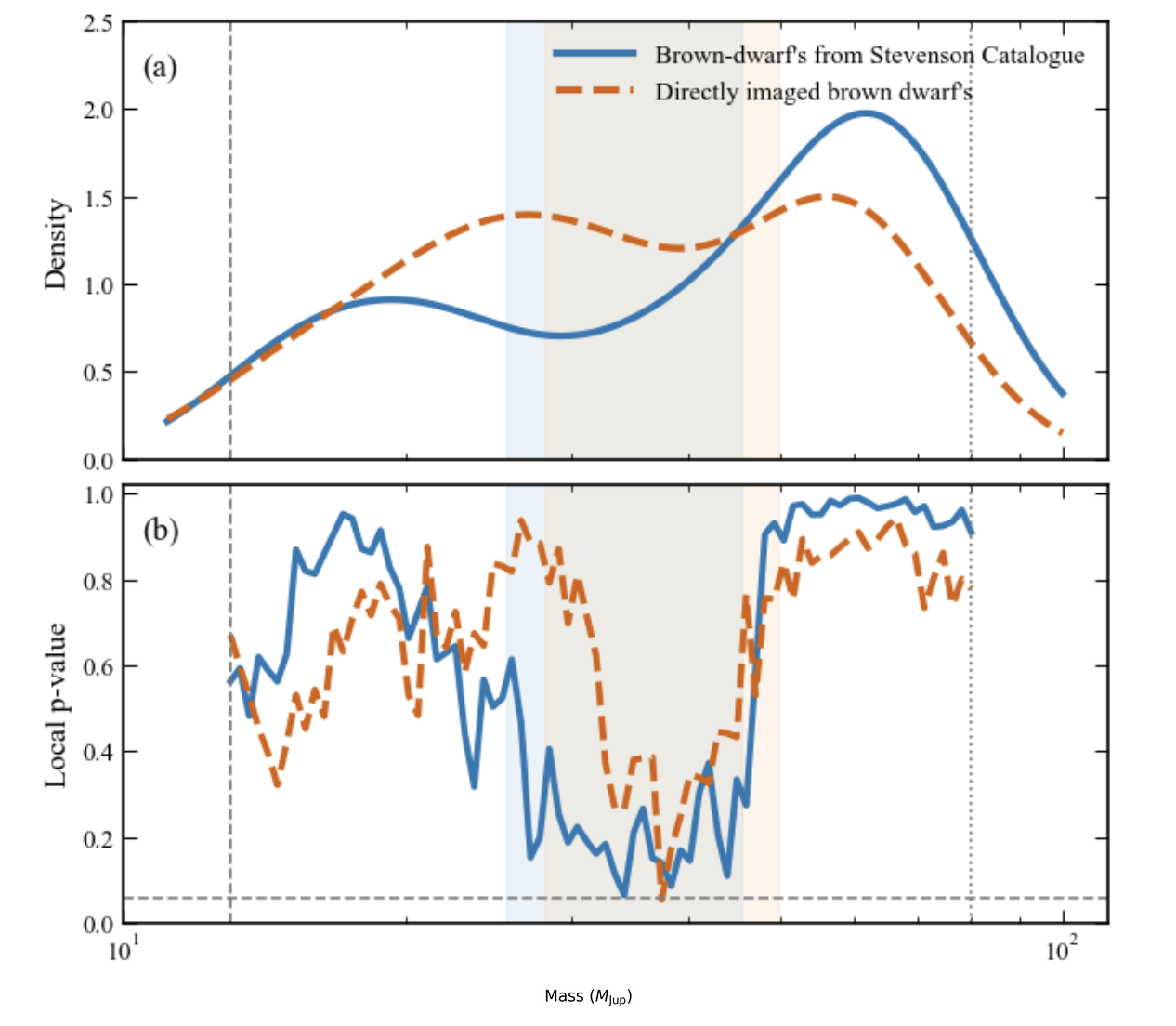}
\caption{Mass distribution of brown-dwarf companions in the
$13$--$80\,M_{\rm Jup}$ range.
\textit{Panel (a)}: Kernel density estimates for the close-in \citet{Stevenson2023}
sample (blue) and the DIBDs (dashed orange);
shaded regions mark each sample's lowest-local-$p$ interval
($\sim22$--$31\,M_{\rm Jup}$ close-in, $\sim35$--$50\,M_{\rm Jup}$ DIBD).
\textit{Panel (b)}: Local $p$-values from the sliding-window scan
(Sect.~\ref{sec:disc_desert}). The dashed gray line marks $p=0.05$.
After look-elsewhere correction, the global $p$-values are
$\simeq 0.6$ and $\simeq 0.2$ (Sect.~\ref{sec:disc_desert}), so the
scan alone does not establish a significant deficit, though the lowest
densities in both samples fall in the same $\sim 20$--$50\,M_{\rm Jup}$
window as the known close-orbit desert.}
\label{fig:bd_desert}
\end{figure}

To localize where each sample is sparsest in mass, we ran a
sliding-window depletion scan over $13 \le M_{\rm comp} \le 80\,M_{\rm Jup}$,
using $80$ logarithmic windows of width $0.15$~dex centered at points
uniformly spaced in $\log_{10} M_{\rm comp}$. At each window center we
counted the observed number of companions and compared it to a null
distribution generated by a smoothed bootstrap of the full BD mass
sample: in each of $N_{\rm sim}=5000$ iterations we resampled with
replacement from the observed log-masses and add Gaussian jitter equal
to the Scott bandwidth of a Gaussian kernel-density estimate of the
sample. The local p-value at each window is
$(N_{\rm sim,\,counts \le obs}+1)/(N_{\rm sim}+1)$, and we took the
smallest value across the scan as the local significance of the
strongest depletion. Because the minimum p-value of a multi-window
scan is biased low, we also computed a global p-value via a
look-elsewhere correction \citep[e.g.,][]{Gross2010}: at each outer
iteration we drew a fresh null dataset by the same smoothed bootstrap,
ran the full scan against an independent ensemble of $512$ null
comparison datasets, and defined the global p-value as the fraction of
outer iterations whose minimum local p is at least as extreme as that
observed in the data. We applied the scan to both the close-in
\citet{Stevenson2023} compilation and the DIBD subsample, restricted
to the $13$--$80\,M_{\rm Jup}$ regime (Fig.~\ref{fig:bd_desert}). The
local p-value minima fall at $\sim 22$--$31\,M_{\rm Jup}$ in the
close-in sample (local $p \simeq 0.04$) and at
$\sim 35$--$50\,M_{\rm Jup}$ in the DIBD subsample (local
$p \simeq 0.013$). After the look-elsewhere correction, however, the
corresponding global p-values rise to $\simeq 0.6$ and $\simeq 0.2$,
respectively. Two points follow. First, the desert itself is not in doubt: it is
firmly established at close orbits by the much larger RV and transit
demographics \citep{Grether2006,Ma2014} and appears at the expected
location in our mass--separation plane (Fig.~\ref{mass_sem}). Second,
the scan is simply underpowered at the present sample sizes to confirm
that deficit on its own; the nonsignificant global p-values reflect
the small number of systems, not the absence of a desert. What the
scan does show is that the lowest-density windows in both samples
coincide with the $\sim 20$--$50\,M_{\rm Jup}$ range where the
close-orbit desert is independently known to lie, so it is consistent
with the established picture rather than evidence against it. The
interpretation below rests on the combination of the
established close-orbit deficit and the well-populated wide-orbit
regime in the same mass range, not on the scan in isolation.

The contrast between the close-orbit deficit and the populated
wide-orbit regime in the same mass range can be understood as a
combination of where each formation channel operates and where
post-formation evolution acts. In the $25$--$50\,M_{\rm Jup}$ range,
core accretion struggles to deliver this much gas because gap
opening limits the supply
\citep{Lubow1999,Mordasini2012}, and cloud fragmentation
preferentially forms objects above the Jeans mass, which in typical
molecular-cloud conditions is $\gtrsim30$--$50\,M_{\rm Jup}$
\citep{Bate2002}. Disk gravitational instability can populate the
intermediate range, but fragments form preferentially at
$a \sim 30$--$100$~au where the disk is massive and cool enough to be
gravitationally unstable \citep{Rice2005,Stamatellos2009,Kratter2016}, not at
sub-au separations. The net result is that intermediate-mass
companions are produced where disk instability operates, precisely the
wide-orbit regime where the DIBD sample lives, and are not
produced in situ at close orbits. Post-formation dynamical
evolution further sculpts the close-orbit distribution: tidal
interactions and migration preferentially remove close-in brown
dwarfs from short-period orbits \citep{Damiani2016}, deepening the
close-orbit desert relative to the underlying formation rate.

\subsection{Caveats and outlook}
\label{sec:disc_caveats}

Several limitations should be borne in mind. The DIBD spectroscopic
sample is modest ($N=31$) and drawn from heterogeneous discovery
programs, which limits our power to resolve subpopulation
differences below $\sim0.1$~dex. Most companion masses are derived
from substellar evolutionary models tied to host-star ages and carry
systematic uncertainties of $20$--$30\%$; where dynamical masses are
available they can
differ from model values by several sigma
\citep[e.g.,][]{Brandt2021}. Direct-imaging surveys preferentially
target young, nearby stars, complicating direct comparisons with the
close-in RV and transit populations, and for most companions only
projected separations are known, limiting period- and
eccentricity-based diagnostics. In addition, the inclusive $80~M_{\rm Jup}$ selection bound lies above the hydrogen-burning limit ($\sim73~M_{\rm Jup}$), so a few of the most massive companions may be low-mass stars rather than brown dwarfs. These limitations will ease over the
coming decade as \emph{Gaia}~DR4 and the next generation of
high-contrast imagers (VLT/ERIS, JWST/NIRCam, and ELT/METIS) expand the
dynamically constrained sample and reach lower companion masses
around mature field stars.

\section{Conclusions}
\label{sec:conclusions}

We compiled 54 directly imaged companions selected over $13$--$80~M_{\rm Jup}$ and homogeneously derived atmospheric parameters and metallicities for the 31 hosts with suitable archival HARPS, UVES, FEROS, or HIRES spectra. Companion masses and projected separations are adopted from the literature. Galactic space velocities are available for 46 hosts and are used only to characterize the selection function of direct imaging. The principal results are as follows.

\begin{enumerate}
\item The DIBD hosts span $-0.56\leq{\rm [Fe/H]}\leq+0.33$~dex, with median $+0.06$~dex and MAD $0.11$~dex. We detect no statistically significant difference between the DIBD metallicity distribution and those of the close-in brown-dwarf or DIP hosts at the current sample sizes. Only the SP--GP and GP--BD-other comparisons remain significant at $p<0.001$ after Bonferroni correction; the GP--DIBD comparison is not significant.

\item Splitting the DIBD sample at $40~M_{\rm Jup}$ yields median metallicities of $+0.04$ and $+0.08$~dex. The present data show no statistically significant evidence of a sharp metallicity transition within the brown-dwarf regime, although the modest sample size limits sensitivity to small offsets.

\item The DIBD hosts are kinematically cold (median $v_{\rm pec}\sim13$~km\,s$^{-1}$ and $\sigma_v\sim23$~km\,s$^{-1}$) and concentrated in the thin disk. The DIP hosts show similarly cold kinematics. This is expected from the preference of imaging surveys for young, nearby targets and is not an independent constraint on formation.

\item The $\sim35$--$55~M_{\rm Jup}$ deficit seen at close orbits is not apparent in the wide-orbit imaging sample. This contrast is consistent with the brown-dwarf desert being primarily a close-orbit demographic feature rather than a global absence of companions in this mass interval.
\end{enumerate}

The absence of a strong GP-like metal-rich bias suggests that classical core accretion is unlikely to dominate the wide-orbit brown-dwarf population. Metallicity-insensitive channels such as disk instability and cloud fragmentation remain plausible, but the present metallicity data and the heuristic mass--separation assignments cannot distinguish these channels for individual systems. These conclusions are population-level and exploratory: the spectroscopic sample contains only 31 hosts, the reference samples are highly unbalanced, the BD-other metallicities are heterogeneous, and objects near the $70$--$75~M_{\rm Jup}$ hydrogen-burning transition may be very low-mass stars. Larger homogeneous samples and improved dynamical masses from future \emph{Gaia} releases and high-contrast follow-up will provide stronger tests.

\begin{acknowledgements}
We thank the anonymous referee for a careful and constructive report that improved this paper. C.S. and S.F. are funded by the European Union (ERC, UNVEIL, 101076613). Views and opinions expressed are however those of the authors only and do not necessarily reflect those of the European Union or the European Research Council. Neither the European Union nor the granting authority can be held responsible for them. C.S. acknowledges financial contribution from PRIN-MUR 2022YP5ACE. This project received funding from the Department of Science and Technology, India as PhD fellowship
of C. Swastik. This research was supported in part by the International Centre for Theoretical Sciences (ICTS) for participating in the programme Exploring Exoplanets: The Journey to Discover New Frontiers and Habitable Worlds 2026 (code: ICTS/EXOPLANETS2026/03). This research made use of the SIMBAD database and the VizieR catalogue access tool, CDS, Strasbourg, France. Computational resources were provided by the INDACO Platform, a High Performance Computing project at the University of Milan, and by CINECA (\url{https://www.unimi.it}, \url{https://www.cineca.it}). We gratefully acknowledge CINECA for providing computational resources and technical support. We also thank the staff of the Indian Astronomical Observatory (IAO), Hanle, for their assistance and support.
C.S. fondly remembers Lalu, a dear companion who is now in the land of the unknown and whose presence brightened the months during which this work was carried out.
\end{acknowledgements}

\bibliographystyle{aa}
\bibliography{biblio}

\onecolumn
\begin{appendix}

\section{Heuristic formation-channel classification}
\label{app:formation}

Table~\ref{tab:targets_formation} gives an illustrative classification based on the literature mass ratio $q=M_{\rm comp}/M_\star$ and projected separation $s$. The assignments are explicitly model-dependent and are not new measurements of this work. Host metallicity is used only as supporting information and never as a decisive object-by-object criterion. Migration, scattering, and mass uncertainties can move systems across the overlapping regimes in Table~\ref{tab:formation_mechanism}; consequently, only the population-level distribution should be interpreted. Entries marked with superscript (a) are post-migration disk-instability or cloud-fragmentation candidates; for HD~206893 and HIP~21152 a core-accretion or boundary origin cannot be excluded. HD~19467~B, HR~7672~B, and HD~4747~B overlap the model-dependent hydrogen-burning transition and are labeled as boundary objects without an assigned formation pathway.

\begin{table*}[h!]
\centering
\tiny
\caption{Heuristic formation-channel assignments for the spectroscopically characterized directly imaged companions.}
\label{tab:targets_formation}
\setlength{\tabcolsep}{3pt}
\begin{tabular}{lcccccl>{\raggedright\arraybackslash}p{0.18\textwidth}}
\hline\hline
Star & $M_\star$ ($M_\odot$) & $M_{\rm comp}$ ($M_{\rm Jup}$) & $s$ (au) & $q$ & [Fe/H] (dex) & Mass reference & Indicative pathway \\
\hline
HD 984 & 1.20 & 61 & 28 & 0.049 & $+0.10$ & \citealt{Franson2022} & Disk instability \\
HD 19467 & 0.96 & 71.6 & 46.9 & 0.071 & $-0.18$ & \citealt{Crepp2014,Maire2020} & Boundary object; pathway not assigned \\
HIP 74865 & 1.42 & 28 & 23 & 0.019 & $+0.07$ & \citealt{Hinkley2015} & Disk instability \\
PZ Tel & 1.10 & 28 & 27 & 0.024 & $0.00$ & \citealt{Franson2023PZTel} & Disk instability \\
GSC 08047$-$00232 & 0.85 & 25 & 264 & 0.028 & $+0.03$ & \citealt{Wahhaj2011} & Cloud fragmentation \\
GJ 229 & 0.58 & 71.4 & 42.9 & 0.118 & $+0.17$ & \citealt{Brandt2020,Xuan2024Natur} & Cloud fragmentation \\
CD$-$35 2722 & 0.55 & 31 & 204 & 0.054 & $-0.15$ & \citealt{Wahhaj2011} & Cloud fragmentation \\
HD 206893 & 1.32 & 17.5 & 8.93 & 0.013 & $+0.08$ & \citealt{Delorme2017} & Disk instability\tablefootmark{a} \\
G 196-3 & 0.40 & 15 & 300 & 0.036 & $+0.03$ & \citealt{Rebolo1998} & Cloud fragmentation \\
GJ 758 & 0.97 & 37.9 & 27.4 & 0.037 & $+0.06$ & \citealt{Bowler2018Gl758} & Disk instability \\
HD 3651 & 0.88 & 53 & 480 & 0.058 & $+0.08$ & \citealt{Bowler2018Gl758} & Cloud fragmentation \\
HD 97334 & 1.05 & 51.5 & 1970 & 0.047 & $+0.20$ & \citealt{DupuyLiu2017} & Cloud fragmentation \\
HII 1348 & 0.85 & 59 & 181 & 0.066 & $+0.33$ & \citealt{Aller2013} & Cloud fragmentation \\
HR 7672 & 1.11 & 75.4 & 19 & 0.065 & $+0.01$ & \citealt{Li2026} & Boundary object; pathway not assigned \\
HD 167665 & 1.01 & 60.3 & 5.5 & 0.057 & $+0.09$ & \citealt{Patel2007} & Disk inst./cloud frag.\tablefootmark{a} \\
HD 33632 & 1.10 & 51.7 & 24.1 & 0.045 & $-0.30$ & \citealt{Currie2020} & Disk instability \\
HD 49197 & 1.10 & 63.2 & 29.1 & 0.055 & $-0.11$ & \citealt{MetchevHillenbrand2005} & Disk inst./cloud frag. \\
HD 72946 & 0.99 & 69.5 & 6.45 & 0.067 & $+0.20$ & \citealt{Bouchy2016} & Disk inst./cloud frag.\tablefootmark{a} \\
HD 13724 & 1.04 & 50.5 & 26.3 & 0.046 & $+0.19$ & \citealt{Rickman2019} & Disk instability \\
HD 176535 & 0.79 & 65.9 & 13 & 0.080 & $-0.03$ & \citealt{Li2023} & Disk inst./cloud frag.\tablefootmark{a} \\
HD 4113 & 1.05 & 36 & 50.4 & 0.033 & $+0.14$ & \citealt{Cheetham2018} & Disk inst./cloud frag. \\
HD 4747 & 0.86 & 70.0 & 10.0 & 0.078 & $-0.20$ & \citealt{Peretti2019} & Boundary object; pathway not assigned \\
HIP 21152 & 1.30 & 24 & 16 & 0.018 & $+0.17$ & \citealt{Kuzuhara2022} & Disk instability\tablefootmark{a} \\
AB Pic & 0.95 & 13.5 & 242 & 0.014 & $+0.04$ & \citealt{Chauvin2005} & Cloud fragmentation \\
HN Peg & 1.10 & 22.0 & 795 & 0.019 & $0.00$ & \citealt{Luhman2007} & Cloud fragmentation \\
HR 2562 & 1.30 & 30 & 22.2 & 0.022 & $+0.21$ & \citealt{Konopacky2016} & Disk instability \\
HD 203030 & 0.95 & 24.1 & 487 & 0.024 & $+0.30$ & \citealt{Metchev2006} & Cloud fragmentation \\
CT Cha & 0.55 & 17 & 500 & 0.030 & $-0.56$ & \citealt{Schmidt2008} & Cloud fragmentation \\
GQ Lup & 0.70 & 20 & 93 & 0.027 & $-0.35$ & \citealt{Seperuelo2008} & Disk inst./cloud frag. \\
ROXs 12 & 0.65 & 16 & 210 & 0.024 & $+0.14$ & \citealt{Bowler2017} & Cloud fragmentation \\
GSC 06214$-$00210 & 0.60 & 16 & 320 & 0.026 & $-0.06$ & \citealt{Bowler2011} & Cloud fragmentation \\
\hline
\end{tabular}
\tablefoot{The values of $M_\star$, $M_{\rm comp}$, and $s$ are adopted from the cited literature and remain heterogeneous and, in many cases, model-dependent. \tablefoottext{a}{Post-migration disk-instability or cloud-fragmentation candidates.}}
\end{table*}

\clearpage
\section{Gaia DR3 astrometric parameters}
\label{app:gaia}

Table~\ref{tab:gaia_params_gt13mj} lists the \emph{Gaia}~DR3 astrometric and RV parameters used in the kinematic analysis for all 54 unique host stars.

\begin{table*}[ht]
\centering
\tiny
\caption{\emph{Gaia} DR3 parameters for the 54 unique brown-dwarf host stars.}
\label{tab:gaia_params_gt13mj}
\setlength{\tabcolsep}{3pt}
\begin{tabular}{l c c c c c c c}
\hline\hline
Name & \emph{Gaia} DR3 & $\alpha$ & $\delta$ & $\varpi$ & $\mu_{\alpha}\cos\delta$ & $\mu_{\delta}$ & $\gamma$ \\
 &  & (deg) & (deg) & (mas) & (mas yr$^{-1}$) & (mas yr$^{-1}$) & (km s$^{-1}$) \\
\hline
HD 984 & 2431157720981843200 & $3.543190$ & $-7.199417$ & $21.8770 \pm 0.0249$ & $104.775 \pm 0.036$ & $-68.016 \pm 0.022$ & $1.038 \pm 0.254$ \\
HD 19467 & 5155981730587168384 & $46.827356$ & $-13.762941$ & $31.2191 \pm 0.0240$ & $-8.694 \pm 0.027$ & $-260.642 \pm 0.026$ & $7.006 \pm 0.127$ \\
HD 176535 & 4101624159180751104 & $285.331099$ & $-13.690821$ & $27.0325 \pm 0.0178$ & $-24.927 \pm 0.018$ & $-29.707 \pm 0.016$ & $-32.187 \pm 0.182$ \\
HD 167665 & 4052127925286198400 & $274.349667$ & $-28.289624$ & $32.3999 \pm 0.0903$ & $133.261 \pm 0.094$ & $-151.806 \pm 0.065$ & $8.407 \pm 0.226$ \\
HD 33632 & 187247281185703296 & $78.321889$ & $37.336703$ & $37.8953 \pm 0.0263$ & $-144.922 \pm 0.031$ & $-136.772 \pm 0.022$ & $-1.754 \pm 0.124$ \\
HD 3651 & 2802397960855105920 & $9.838653$ & $21.248833$ & $90.0248 \pm 0.0482$ & $-461.948 \pm 0.068$ & $-369.624 \pm 0.025$ & $-33.065 \pm 0.126$ \\
HD 4113 & 5000774703569900800 & $10.802760$ & $-37.983141$ & $23.8256 \pm 0.0240$ & $49.412 \pm 0.016$ & $-114.290 \pm 0.024$ & $5.051 \pm 0.121$ \\
HD 4747 & 2348830516542653824 & $12.364012$ & $-23.211912$ & $53.0526 \pm 0.0282$ & $519.049 \pm 0.033$ & $124.041 \pm 0.032$ & $9.708 \pm 0.119$ \\
HD 49197 & 952346742338146176 & $102.338667$ & $43.758879$ & $24.1155 \pm 0.0318$ & $-37.161 \pm 0.029$ & $-50.941 \pm 0.024$ & $10.507 \pm 0.251$ \\
HD 72946 & 594989417312890368 & $128.963000$ & $6.622156$ & $38.9809 \pm 0.0412$ & $-136.593 \pm 0.044$ & $-137.148 \pm 0.034$ & $28.705 \pm 0.120$ \\
HD 206893 & 6843672087120107264 & $326.341701$ & $-12.783352$ & $24.5275 \pm 0.0354$ & $94.112 \pm 0.035$ & $-0.463 \pm 0.026$ & $-11.799 \pm 0.140$ \\
HD 130948 & 1265976524286377856 & $222.566581$ & $23.911984$ & $54.9502 \pm 0.0343$ & $144.395 \pm 0.024$ & $31.661 \pm 0.037$ & $-2.626 \pm 0.122$ \\
HD 13724 & 4942867480584948352 & $33.085956$ & $-46.816683$ & $23.0157 \pm 0.0178$ & $-30.624 \pm 0.016$ & $-69.088 \pm 0.020$ & $20.384 \pm 0.124$ \\
HII 1348 & 66734720809017856 & $56.825356$ & $24.390565$ & $6.9790 \pm 0.0305$ & $21.257 \pm 0.042$ & $-45.555 \pm 0.027$ & $6.428 \pm 2.963$ \\
HIP 64892 & 6088264477373916032 & $199.521078$ & $-44.055473$ & $8.3595 \pm 0.0483$ & $-30.729 \pm 0.039$ & $-20.073 \pm 0.034$ & \ldots \\
HIP 78530 & 6243537406369453824 & $240.481003$ & $-21.980495$ & $7.4238 \pm 0.0299$ & $-12.114 \pm 0.033$ & $-24.040 \pm 0.022$ & $-4.688 \pm 0.860$ \\
HIP 79098 & 6242058872466158848 & $242.182119$ & $-23.685528$ & $6.4762 \pm 0.1839$ & $-10.087 \pm 0.224$ & $-26.085 \pm 0.179$ & \ldots \\
HIP 79124 & 6248783275121192704 & $242.260802$ & $-18.995674$ & $7.3721 \pm 0.0296$ & $-7.551 \pm 0.038$ & $-24.199 \pm 0.025$ & $-8.909 \pm 0.967$ \\
HIP 81208 & 6020514769906985728 & $248.807611$ & $-35.724761$ & $6.8424 \pm 0.0475$ & $-9.701 \pm 0.052$ & $-25.913 \pm 0.040$ & \ldots \\
HIP 79797 & 5821125860988362752 & $244.271999$ & $-67.941659$ & $18.1859 \pm 0.0315$ & $-46.084 \pm 0.016$ & $-84.330 \pm 0.025$ & $-9.768 \pm 0.238$ \\
HIP 21152 & 3285426613077584384 & $68.020537$ & $5.410076$ & $23.1089 \pm 0.0278$ & $112.174 \pm 0.035$ & $7.756 \pm 0.022$ & $41.031 \pm 0.137$ \\
HN Peg & 1772187382746856320 & $326.131604$ & $14.771437$ & $55.1480 \pm 0.0348$ & $231.108 \pm 0.030$ & $-113.200 \pm 0.027$ & $-16.938 \pm 0.118$ \\
$\beta$ Cir & 5877059048308526720 & $229.377706$ & $-58.801812$ & $33.8205 \pm 0.2515$ & $-97.182 \pm 0.217$ & $-136.055 \pm 0.238$ & $10.659 \pm 0.153$ \\
CD$-$35 2722 & 2885863400349980288 & $92.330014$ & $-35.825545$ & $44.7203 \pm 0.0128$ & $-3.606 \pm 0.014$ & $-56.078 \pm 0.015$ & $31.211 \pm 0.681$ \\
G 196-3 & 824017070904063104 & $151.088445$ & $50.386153$ & $45.8541 \pm 0.0188$ & $-141.079 \pm 0.016$ & $-202.336 \pm 0.016$ & $-1.592 \pm 0.674$ \\
GJ 229 & 2940856402123426176 & $92.643579$ & $-21.867823$ & $173.5740 \pm 0.0170$ & $-135.692 \pm 0.011$ & $-719.178 \pm 0.017$ & $4.226 \pm 0.119$ \\
GJ 758 & 2046207670631964928 & $290.892157$ & $33.222678$ & $64.0703 \pm 0.0154$ & $81.966 \pm 0.013$ & $160.159 \pm 0.016$ & $-21.623 \pm 0.118$ \\
GSC 06214$-$00210 & 6244440552088537600 & $245.477689$ & $-20.719344$ & $9.1923 \pm 0.0214$ & $-19.542 \pm 0.027$ & $-31.144 \pm 0.023$ & $-6.290 \pm 0.670$ \\
GSC 08047$-$00232 & 4913138232358905216 & $28.061304$ & $-52.325931$ & $11.5431 \pm 0.0122$ & $50.045 \pm 0.011$ & $-10.474 \pm 0.012$ & $13.350 \pm 1.247$ \\
HR 2562 & 5480006192386819456 & $102.504273$ & $-60.248662$ & $29.4738 \pm 0.0185$ & $4.830 \pm 0.025$ & $108.527 \pm 0.023$ & $26.186 \pm 0.158$ \\
HR 3549 & 5304593027881569024 & $133.265565$ & $-56.649323$ & $10.5509 \pm 0.0383$ & $-22.406 \pm 0.041$ & $36.523 \pm 0.038$ & $31.275 \pm 0.643$ \\
HR 7329 & 6643602576214115584 & $290.713556$ & $-54.424298$ & $20.6028 \pm 0.0988$ & $25.824 \pm 0.073$ & $-82.965 \pm 0.061$ & $3.798 \pm 0.592$ \\
HR 7672 & 1821708351374312064 & $301.024118$ & $17.068324$ & $56.2693 \pm 0.0433$ & $-387.472 \pm 0.037$ & $-419.497 \pm 0.029$ & $5.140 \pm 0.124$ \\
PZ Tel & 6655168686921108864 & $283.274586$ & $-50.180907$ & $21.1621 \pm 0.0223$ & $16.273 \pm 0.018$ & $-85.519 \pm 0.017$ & $-3.595 \pm 1.546$ \\
ROXs 12 & 6048935358761628288 & $246.616798$ & $-25.446698$ & $7.2170 \pm 0.0172$ & $-6.858 \pm 0.020$ & $-24.815 \pm 0.013$ & $-8.001 \pm 5.362$ \\
TWA 5 & 5398663566249861120 & $172.979796$ & $-34.607657$ & $20.1311 \pm 0.0572$ & $-84.789 \pm 0.063$ & $-22.544 \pm 0.049$ & $5.872 \pm 1.976$ \\
HR 7871 & 1804510069668710400 & $308.827443$ & $14.674266$ & $14.9441 \pm 0.1284$ & $45.504 \pm 0.154$ & $11.510 \pm 0.121$ & $-27.680 \pm 0.368$ \\
HD 284149 & 51914888911355776 & $61.661690$ & $20.303032$ & $8.5118 \pm 0.0181$ & $6.504 \pm 0.022$ & $-14.715 \pm 0.017$ & $14.034 \pm 0.713$ \\
GJ 1048 & 5125414998097353600 & $39.003636$ & $-23.521262$ & $47.0903 \pm 0.0223$ & $83.275 \pm 0.019$ & $14.006 \pm 0.021$ & $16.251 \pm 0.123$ \\
GJ 570 & 6232511606838403968 & $224.371594$ & $-21.423140$ & $169.8843 \pm 0.0653$ & $1031.472 \pm 0.068$ & $-1723.619 \pm 0.055$ & $26.751 \pm 0.118$ \\
HD 126053 & 3654496279558010624 & $215.814680$ & $1.239441$ & $57.2706 \pm 0.0375$ & $223.531 \pm 0.048$ & $-478.275 \pm 0.031$ & $-19.209 \pm 0.137$ \\
HIP 70849 & 6099000589942631424 & $217.327067$ & $-46.464712$ & $41.4618 \pm 0.0175$ & $-44.051 \pm 0.017$ & $-201.577 \pm 0.020$ & $-0.105 \pm 0.162$ \\
HIP 74865 & 6211094559143445888 & $229.483686$ & $-30.478299$ & $8.0898 \pm 0.0190$ & $-21.239 \pm 0.021$ & $-28.373 \pm 0.018$ & $-0.145 \pm 0.845$ \\
LSPM J1446$+$4633 & 1590096093840214272 & $221.506111$ & $46.555032$ & $58.6329 \pm 0.1135$ & $496.880 \pm 0.107$ & $-415.588 \pm 0.121$ & $13.840 \pm 0.923$ \\
Gliese 569 & 1187851653287128576 & $223.623109$ & $16.100531$ & $100.5243 \pm 0.0210$ & $279.117 \pm 0.022$ & $-117.908 \pm 0.021$ & $-7.966 \pm 0.175$ \\
Gliese 337 & 607464716759460480 & $138.070706$ & $14.997134$ & \ldots & \ldots & \ldots & $52.547 \pm 6.220$ \\
HD 203030 & 1846882224145757056 & $319.743245$ & $26.230586$ & $25.4595 \pm 0.0212$ & $133.810 \pm 0.021$ & $9.245 \pm 0.018$ & $-16.934 \pm 0.146$ \\
HIP 38939 & 5698188160215537920 & $119.519996$ & $-25.627721$ & $54.1565 \pm 0.0183$ & $362.409 \pm 0.012$ & $-245.786 \pm 0.016$ & $-8.193 \pm 0.119$ \\
CT Cha & 5201360671411974912 & $166.037912$ & $-76.455369$ & $5.2645 \pm 0.0116$ & $-22.223 \pm 0.014$ & $0.019 \pm 0.013$ & \ldots \\
GQ Lup & 6011522757643074304 & $237.300362$ & $-35.651509$ & $6.4893 \pm 0.0289$ & $-14.133 \pm 0.032$ & $-23.329 \pm 0.026$ & \ldots \\
AB Pic & 5495052596695570816 & $94.803925$ & $-58.054112$ & $19.9452 \pm 0.0124$ & $14.314 \pm 0.015$ & $45.234 \pm 0.017$ & $22.022 \pm 0.190$ \\
$\kappa$ And & 1926476042681491712 & $355.102613$ & $44.333848$ & $19.4064 \pm 0.2104$ & $79.998 \pm 0.156$ & $-19.011 \pm 0.128$ & \ldots \\
HD 97334 & 761919883981626752 & $168.133427$ & $35.813407$ & $44.0583 \pm 0.0232$ & $-249.194 \pm 0.018$ & $-151.406 \pm 0.026$ & $-3.794 \pm 0.128$ \\
HD 151985 & 5971244451311982336 & $253.083870$ & $-38.017637$ & $5.6632 \pm 0.2750$ & $-12.114 \pm 0.300$ & $-22.570 \pm 0.266$ & \ldots \\
\hline
\end{tabular}
\tablefoot{Coordinates are given for epoch 2016.0 (ICRS). Ellipses indicate parameters not available in \emph{Gaia}~DR3: Gliese~337 lacks the astrometry required for the space-velocity computation, and seven hosts lack a \emph{Gaia}~DR3 RV.}
\end{table*}

\end{appendix}

\end{document}